\documentclass[fleqn,usenatbib]{mnras}

\usepackage[normalem]{ulem}
\usepackage[T1]{fontenc}
\usepackage{ae,aecompl}
\usepackage{enumitem}
\usepackage{multirow}
\usepackage{tabularx}

\DeclareRobustCommand{\VAN}[3]{#2}
\let\VANthebibliography\thebibliography
\def\thebibliography{\DeclareRobustCommand{\VAN}[3]{##3}\VANthebibliography}

\usepackage{graphicx}	
\usepackage{amsmath}	
\usepackage{amssymb}	

\newcommand{\hi}{H\textsc{i}\ }
\newcommand{\hinospace}{\textrm{H\textsc{i}}}
\newcommand{\size}[2]{{\fontsize{#1}{0}\selectfont#2}}

\newcommand{\fastica}{\size{7.5}{FASTICA}}
\newcommand{\multidark}{\textsc{MultiDark}}
\newcommand{\deltadiff}{\delta\hspace{-0.3mm}}
\newcommand{\diff}{\text{d}\hspace{-0.25mm}}
\newcommand{\secref}[1]{\hyperref[#1]{Section~\ref*{#1}}}
\newcommand{\appref}[1]{\hyperref[#1]{Appendix~\ref*{#1}}}

\title[HI IM: unbiased parameter estimation]{Power spectrum multipole expansion for HI intensity mapping experiments: unbiased parameter estimation}

\author[P. S. Soares et al.]{Paula S. Soares$^{1}$\thanks{E-mail: p.s.soares@qmul.ac.uk},
Steven Cunnington$^{1}$,
Alkistis Pourtsidou$^{1,2}$,
and Chris Blake$^{3}$
\newauthor
\\
$^{1}$School of Physics and Astronomy, Queen Mary University of London, Mile End Road, London E1 4NS, UK\\
$^{2}$Department of Physics \& Astronomy, University of the Western Cape, Cape Town 7535, South Africa\\
$^{3}$Centre for Astrophysics \& Supercomputing, Swinburne University of Technology, P.O. Box 218, Hawthorn, VIC 3122, Australia
}

\date{Accepted XXX. Received YYY; in original form ZZZ}

\pubyear{2020}

\begin{document}
\label{firstpage}
\pagerange{\pageref{firstpage}--\pageref{lastpage}}
\maketitle

\begin{abstract}
We assess the performance of the multipole expansion formalism in the case of single-dish \hi intensity mapping, including instrumental and foreground removal effects. This formalism is used to provide MCMC forecasts for a range of \hi and cosmological parameters, including redshift space distortions and the Alcock-Paczynski effect. We first determine the range of validity of our power spectrum modelling by fitting to simulation data, concentrating on the monopole, quadrupole, and hexadecapole contributions. 
We then show that foreground subtraction effects can lead to severe biases in the determination of cosmological parameters, in particular the parameters relating to the transverse BAO rescaling, the growth rate and the \hi bias ($\alpha_\perp$, $\overline{T}_\hinospace f\sigma_8$, and $\overline{T}_\hinospace b_\hinospace \sigma_8$, respectively). We attempt to account for these biases by constructing a 2-parameter foreground modelling prescription, and find that our prescription  leads to unbiased parameter estimation at the expense of increasing the estimated uncertainties on cosmological parameters. In addition, we confirm that instrumental and foreground removal effects significantly impact the theoretical covariance matrix, and cause the covariance between different multipoles to become non-negligible. Finally, we show the effect of including higher-order multipoles in our analysis, and how these can be used to investigate the presence of instrumental and systematic effects in \hi intensity mapping data.
\end{abstract}




\section{Introduction}

The standard cosmological model, $\Lambda$CDM, describes a universe with zero spatial curvature, containing cold dark matter and dark energy in the form of a cosmological constant ($\Lambda$), which drives late time cosmic acceleration. It has 6 parameters, 5 of which have been measured to within 1\% precision through observations of the Cosmic Microwave Background \citep{planckcollaboration2018}. Large Scale Strucrure (LSS) surveys have also provided observations that are in very good agreement with $\Lambda$CDM \citep{Anderson_2014, Song_2016, Alam_2017, Beutler_2016, Tr_ster_2020, ebosscollaboration2020}. LSS surveys in particular are able to probe whether general relativity is the correct description of gravity on cosmological scales by measuring the logarithmic growth rate of structure ($f$)  \citep{Guzzo_2008}. This parameter can be measured through the redshift space distortion (RSD) signature on the 2-point statistics of galaxy clustering \citep{Blake_2011, Reid_2012, Macaulay_2013, Beutler_2014, Gil_Mar_n_2016, Simpson_2016, Icaza_Lizaola_2019}.

Neutral hydrogen (\hinospace) Intensity Mapping (IM) is a novel technique that is able to efficiently and rapidly observe a very wide redshift range, including high redshifts, $z>3$, that are inaccessible by current and forthcoming optical galaxy surveys (see \citet{kovetz2017lineintensity} for a review). In particular, \hi IM treats the 21cm sky as a diffuse background and measures its intensity in large voxels, as opposed to detecting individual galaxies \citep{Battye2004, Chang_2008, Wyithe_2009, Mao_2008, peterson2009, Seo_2010, Ansari_2012}. In the post-reionization Universe, neutral hydrogen resides inside galaxies where it is self-shielded from ionization; it can thus be used as a tracer of the underlying matter distribution.  Using \hi IM, it is possible to map the 3D LSS of the Universe, and probe the underlying cosmology through the \hi power spectrum.

Current \hi IM detections come from cross-correlating \hi IM maps from the Green Bank Telescope (GBT) or the Parkes radio telescope with optical galaxy surveys, probing the clustering of neutral hydrogen at $z<1$ \citep{chang2010, Masui:2012zc, Switzer_2013, Wolz_2016, Anderson_2018, Li:2020pre}. More specifically, the GBT has constrained the combination of the \hi abundance ($\Omega_\hinospace$) and linear \hi bias ($b_\hinospace$) at $z=0.8$, $\Omega_\hinospace b_\hinospace r = [4.3 \pm 1.1] \times 10^{-4}$, using cross-correlation with the WiggleZ optical galaxy survey, where $r$ is the galaxy-hydrogen correlation coefficient \citep{Masui:2012zc}. A detection through auto-correlation is yet to be made due to residual systematic effects. However, since most of these systematic effects do not correlate with  optical galaxy surveys, they are mitigated in cross-correlation. 

The Square Kilometre Array (SKA)\footnote{\href{www.skatelescope.org}{www.skatelescope.org}} will be a radio observatory able to reach unprecedented statistical precision on \hi IM measurements (see e.g. \citet{santos2015cosmology},  \citet{Bacon:2018dui}), assuming that systematic effects are controlled or mitigated. In the case of \hi IM, large galactic and extragalactic foregrounds dominate over the signal by several orders ot magnitude. However, in principle we can differentiate these dominant foregrounds from the signal since they are expected to be smooth in frequency \citep{Liu_2011, chang2010, Wolz_2014, Alonso_2014, Bigot_Sazy_2015, Olivari_2015, Switzer_2015, wolz2015foreground,Cunnington:2020njn}.

In addition, it is not yet fully understood how the systematic effects present in an \hi IM survey affect its noise properties, in particular the covariance matrix. The effect of foreground removal has been shown to mostly affect power on large scales. The telescope beam significantly damps power on scales smaller than its resolution, but it also affects larger scales \citep{VillaescusaNavarro:2016bao,Cunnington:2020mnn}.  Analytically, both of these effects carry over to the theoretical covariance matrix in the case of the \hi IM power spectrum \citep{Bernal:2019lim}.

In this paper, we build upon the work of \citet{Cunnington:2020mnn} (hereafter C20) and \citet{Blake:2019ddd}. \citet{Blake:2019ddd} studied the modelling of the \hi IM power spectrum including observational effects, and C20 extended this into a comprehensive simulations and data analysis pipeline\footnote{\href{https://github.com/IntensityTools/MultipoleExpansion}{github.com/IntensityTools/MultipoleExpansion}} for analysing the \hi IM power spectrum multipoles taking into account instrumental and foreground removal effects. Here we extend on C20 to perform cosmological parameter estimation with the \hi IM power spectrum, using simulations that include the relevant instrumental and foreground removal effects. In particular, we use Markov Chain Monte Carlo (MCMC) analyses to forecast uncertainties for a range of \hi and cosmological parameters. We aim to realistically assess what a future SKA-like \hi IM survey will be able to constrain, and in particular how foreground removal affects cosmological parameter estimation. We are interested in both precision and accuracy, i.e. we pay particular attention to the requirement of \emph{unbiased} parameter estimation.

The paper is structured as follows: In \secref{sec:model}, we describe the observed \hi IM power spectrum, including modelling of the telescope beam and foreground removal, decomposed into multipoles. In \secref{sec:sim}, we describe our IM simulations. In \secref{sec:results}, we test the range of validity of our model using the simulations, report our MCMC analysis results, look into how instrumental and systematic effects affect the noise covariance matrix, and investigate whether higher order multipoles can add useful information. We conclude in \secref{sec:conclusion}. 

Throughout this paper, we assume a flat $\Lambda\text{CDM}$ cosmology consistent with the \textsc{Planck}15 analysis \citep{Ade:2015xua}, with $\Omega_\text{M} = 0.307$, $\Omega_\text{b} = 0.048$, $\Omega_\Lambda = 0.693$,  $\sigma_8 = 0.823$, $n_\text{s} = 0.96$ and Hubble parameter $h=0.678$.

\section{Model}\label{sec:model}

\subsection{Redshift space distortions}

Redshift space distortions (RSD) introduce anisotropies in the observed \hi power spectrum. In order to account for this, we consider the power spectrum as a function of the directional wave vector $\vec{k}$, which can be decomposed into its module $k$ and the cosine of the angle $\theta$ between the wave vector and the line-of-sight (LoS) component $\mu = \vec{k} \cdot k_{\parallel} \equiv \cos\theta$. We model RSD by considering the Kaiser effect \citep{Kaiser:1987qv}, which is a large-scale effect dependent on the growth rate $f$, and the Fingers-of-God (FoG) effect \citep{Jackson:2008yv}, which is a small-scale non-linear effect that depends on the velocity dispersion of the tracer objects ($\sigma_v$). The anisotropic \hi power spectrum can be written as: 
\begin{equation}\label{RSDPk2}
	P_\hinospace(k, \mu) =  \frac{\left( \overline{T}_\hinospace b_\hinospace + \overline{T}_\hinospace f \mu^{2}\right)^{2} P_\text{M}(k)}{1+\left(k \mu \sigma_\text{v} / H_{0}\right)^{2}} + P_\text{SN} \, ,
\end{equation}
where $P_\text{SN} = \overline{T}_\hinospace^2 (1/\overline{n})$ is the shot noise, $\overline{n}$ is the number density of objects, $P_\text{M}(k)$ is the underlying matter power spectrum, and $\overline{T}_\hinospace$ is the mean \hi brightness temperature, modelled as \citep{Battye:2012tg}:
\begin{equation}\label{TbarModelEq}
    \overline{T}\hspace{-0.5mm}_\hinospace(z) = 180\Omega_{\hinospace}(z)h\frac{(1+z)^2}{H(z)/H_0} \, {\text{mK}} \, .
\end{equation}

\subsection{Alcock-Paczynski effect}

When we measure the \hi power spectrum using intensity mapping, we first measure redshifts and then transform these into distances. In order to do this transformation, we must assume a cosmology. If the assumed cosmology does not match the real one, we get further anisotropies in the power spectrum measurements. This is known as the Alcock-Paczynski (AP) effect \citep{Alcock1979}. In the transverse and radial directions respectively, we model these anisotropies as (see e.g. \citet{Euclid:2019sum, Bernal:2019lim}):
\begin{equation}
\begin{split}
    \ & \alpha_\perp = \frac{D_{A}(z)}{D_{A}(z)^{\rm f}}\, ,\\
    \ & \alpha_\parallel = \frac{H(z)^{\rm f}}{H(z)}\, ,
\end{split}
\end{equation}
where throughout the paper the superscript or subscript `f' refers to the fiducial value, in this case our fiducial, assumed cosmology. Here we note that we follow the notation and method of \citet{Euclid:2019sum} and do not include the degeneracy with the sound horizon at radiation drag ($r_s$) in these factors, as is done when performing a BAO-only analysis. This is because we are performing a full shape analysis. These factors distort the perpendicular and parallel to the LoS wave vectors as: 
\begin{equation}
\begin{split}
   \ & k_\perp = k_\perp^{\rm f}/\alpha_\perp \, , \\
   \ & k_\parallel = k_\parallel^{\rm f}/\alpha_\parallel \, .
\end{split}
\end{equation}
It is useful to define the factor $F_{\rm AP} = \alpha_\parallel / \alpha_\perp$, which helps describe how $k$ and $\mu$ become distorted, and how to recover the true underlying value from the fiducial value:
\begin{equation}
\begin{split}
    \ & k = \frac{k^{\rm f}}{\alpha_\perp} \left[ 1 + (\mu^{\rm f})^2 (F_{\rm AP}^{-2} - 1) \right]^{1/2}\, ,\\
    \ & \mu = \frac{\mu^{\rm f}}{F_{\rm AP}} \left[ 1 + (\mu^{\rm f})^2 (F_{\rm AP}^{-2} - 1) \right]^{-1/2}\, .
\end{split}
\end{equation}
The \hi power spectrum can then be described in terms of this effect as:
\begin{equation}\label{RSDPk3}
	P_\hinospace(k^{\rm f}, \mu^{\rm f}) = \alpha_\parallel^{-1} \alpha_\perp^{-2} \left[ \frac{\left( \overline{T}_\hinospace b_\hinospace + \overline{T}_\hinospace f \mu^{2}\right)^{2} P_\text{M}(k)}{1+\left(k \mu \sigma_\text{v} / H_{0}\right)^{2}} + P_\text{SN} \right] \, .
\end{equation}
\subsection{Telescope beam smoothing effect}\label{sec:beammodel}

The telescope beam introduces one of the main instrumental effects in the case of single-dish intensity mapping experiments. We can model this effect using a damping term dependent on the physical smoothing scale of the beam (see e.g. \citet{Battye:2012tg,VillaescusaNavarro:2016bao, Cunnington:2020mnn}).  Assuming the telescope beam can be modelled as a Gaussian, this is defined as $R_\text{beam} = \sigma_\theta r(z)$, where $\sigma_\theta =  \theta_\text{FWHM} / ( 2\sqrt{2\ln(2)} )$, $\theta_\text{FWHM}$ is the full-width-half-maximum of the beam in radians, and $r(z)$ is the comoving distance to a redshift $z$. The Fourier transform of the telescope beam damping term is:
\begin{equation}\label{BeamDampEq}
	\widetilde{B}_\perp(k,\mu) = \exp\left(\frac{-k^2 R_\text{beam}^2(1-\mu^2)}{2}\right) \, ,
\end{equation}
and the power spectrum becomes:
\begin{equation}\label{RSDPk}
\begin{split}
	P_\hinospace(k^{\rm f}, \mu^{\rm f}) = \ & \frac{\widetilde{B}_\perp^2(k,\mu)}{\alpha_\parallel\, \alpha_\perp^{2}} \\
	\ & \times \left[ \frac{\left( \overline{T}_\hinospace b_\hinospace + \overline{T}_\hinospace f \mu^{2}\right)^{2} P_\text{M}(k)}{1+\left(k \mu \sigma_\text{v} / H_{0}\right)^{2}} + P_\text{SN} \right] \, .
\end{split}
\end{equation}

We should also note that for surveys that are limited in frequency resolution, a similar effect will occur on the small radial scales. In our case, and in general for single dish experiments, the frequency resolution is very good (much better than the angular resolution and radial non-linear dispersion effects) and does not cause any discernible effects in our power spectrum measurements, so we do not include it in our modelling. In cases where this might be relevant, a way to account for this smoothing effect is described in \citet{Blake:2019ddd}. Given the resolution in the radial direction is set by the frequency channel bandwidth $\delta_\nu$, the smoothing effect can be modelled as: 
\begin{equation*}
    \tilde{B}_\parallel(k,\mu) = 
    \frac{\sin{(k\mu s_\parallel/2)}}{k\mu s_\parallel/2} \, , 
\end{equation*}
where $s_\parallel = [c/H(z)](1+z)^2 (\delta_\nu/\nu_{21})$, with $\nu_{21}$ being the rest \hi emission frequency.

\subsection{Re-normalising by $\sigma_8$}

In this work, we calculate the underlying non-linear matter power spectrum given a fiducial, assumed cosmology using the \texttt{python} package \texttt{Nbodykit}\footnote{\href{https://nbodykit.readthedocs.io/en/latest/}{https://nbodykit.readthedocs.io}} \citep{Hand:2017pqn}, which uses the CLASS Boltzmann solver \citep{lesgourgues2011cosmic,Blas:2011rf}, and we choose the Halofit prescription \citep{Takahashi:2012em}. It is useful to parametrise this template calculated matter power spectrum $P_\text{M}(k)$ by $\sigma_8$, which is the RMS of the density fluctuations within a sphere of radius $8\,h^{-1}\text{Mpc}$ (see e.g. \citet{Euclid:2019sum} for a more detailed description):
\begin{equation}
    P_\text{M,8}(k) = \frac{P_\text{M}(k)}{\sigma_8^2} \, .
\end{equation}
Including this, our final power spectrum model becomes:
\begin{equation}\label{FullPknoFG}
\begin{split}
	P_\hinospace(k^{\rm f}, \mu^{\rm f}) = \ & \frac{\widetilde{B}_\perp^2(k,\mu)}{\alpha_\parallel\, \alpha_\perp^{2}} \\ 
	\ & \times \left[ \frac{\left( \overline{T}_\hinospace b_\hinospace \sigma_8 + \overline{T}_\hinospace f \sigma_8 \mu^{2}\right)^{2} P_\text{M,8}(k)}{1+\left(k \mu \sigma_\text{v} / H_{0}\right)^{2}} + P_\text{SN} \right] \, .
\end{split}
\end{equation}
The set of parameter combinations that can be measured using this model is:
\begin{equation}
    \vec{\theta} = \{ \alpha_\parallel,\,\, \alpha_\perp,\,\, \overline{T}_\hinospace f\sigma_8,\,\, \overline{T}_\hinospace b_\hinospace \sigma_8,\,\, \sigma_v,\,\, P_\text{SN}\} \, .
\end{equation}

We note that, in comparison to optical galaxy surveys, we have an additional degeneracy coming from the mean brightness temperature $\overline{T}_\hinospace$, which is proportional to $\Omega_\hinospace$. Previous works using Fisher matrix forecasts (see e.g. \citet{Bull_2015, Pourtsidou_2017}) assume this is a known quantity and keep it fixed, but here we choose to include it since $\Omega_\hinospace$ is quite poorly constrained \citep{Crighton:2015pza}. We also note that, as suggested in \cite{Castorina:2019zho}, this degeneracy can be broken by using information from the non-linear regime of structure formation and perturbation theory modelling, but this would require precise and well-calibrated interferometric observations. In this work we are assuming a survey in single-dish mode \citep{Battye:2012tg, Bacon:2018dui} and we are focusing on the beam and foreground removal effects (with the latter being a major issue for both single dishes and interferometers).
\subsection{Multipole expansion}

We can expand the anisotropic power spectrum $P_\hinospace(k, \mu)$ in terms of Legendre polynomials as
\begin{equation}\label{LegExpansion1}
	P_\hinospace(k, \mu)=\sum_{\ell} P_{\ell}(k) \mathcal{L}_{\ell}(\mu) \, ,
\end{equation}
where $\mathcal{L}_{\ell}(\mu)$ is the $\ell^\text{th}$ Legendre polynomial:
\begin{equation}\label{LegPolys}
\begin{split}
	\ & \mathcal{L}_{0} = 1\,, \quad \mathcal{L}_{2} = \frac{3\mu^2 -1}{2}\,, \quad \mathcal{L}_{4} = \frac{35\mu^4 -30\mu^2 + 3}{8}\,, \\
	\ & \mathcal{L}_{6} = \frac{231\mu^6 - 315\mu^4 + 105\mu^2 - 5}{16} \,.
\end{split}
\end{equation}
Our full model, expanded into power spectrum multipoles, is then given by:
\begin{equation}\label{FullModMult}
\begin{split}
	P_{\ell}(k^{\rm f})= \ & \frac{2 \ell+1}{2} (\alpha_\parallel^{-1}\, \alpha_\perp^{-2}) \int_{-1}^{1} \diff \mu^{\rm f} \, \mathcal{L}_{\ell}(\mu^{\rm f}) \,  \widetilde{B}_\perp^2(k,\mu)  \\ 
	\ & \times \left[ \frac{\left( \overline{T}_\hinospace b_\hinospace \sigma_8 + \overline{T}_\hinospace f \sigma_8 \mu^{2}\right)^{2} P_\text{M,8}(k)}{1+\left(k \mu \sigma_\text{v} / H_{0}\right)^{2}} + P_\text{SN} \right] \, .
\end{split}
\end{equation}
We consider the monopole ($P_{0}$), quadrupole ($P_{2}$), hexadecapole ($P_{4}$) and 64-pole ($P_{6}$) in our analysis. 

\subsection{Modelling the effect of foreground removal}\label{sec:FGmodel}

In order to model the effect of foreground removal on the \hi power spectrum, we introduce a damping term inspired by the survey volume damping function (see e.g. \citet{Bernal:2019lim}). This is given in Fourier space by:
\begin{equation}
\begin{split}
    \widetilde{B}_{\rm vol}(k, \mu) = \ & \left( 1 - \mathrm{exp}\left\{ -\left( \frac{k_\perp}{k^{\rm min}_\perp} \right)^2 \right\} \right) \\
    \ & \times \left( 1 - \mathrm{exp}\left\{ -\left( \frac{k_\parallel}{k^{\rm min}_\parallel} \right)^2 \right\} \right)\, ,
\end{split}
\end{equation}
which describes how we are not able to access modes smaller than $k^{\rm min}_\perp$ or $k^{\rm min}_\parallel$ in the perpendicular and parallel to the LoS directions. If we assume a survey box to have comoving distance dimensions given by [$L_x, L_y, L_z$], we have that the smallest (largest) modes (physical scales) accessible in the perpendicular and parallel to the LoS directions are: $k^{\rm min}_\perp = 2\pi/\sqrt{L_x^2 + L_y^2}$ and $k^{\rm min}_\parallel = 2\pi / L_z$.

We assume that the process of foreground removal similarly removes power from modes along the parallel and perpendicular to the LoS directions based on the foreground properties and survey geometry. In this case, we are particularly considering the effects of an Independent Component Analysis (ICA) foreground removal technique (see e.g. \citet{Alonso_2014}). This component separation technique does not try to assume a specific form for the foreground contamination, but relies on the fact that the sources of the foregrounds are statistically independent and can be isolated from the cosmological signal. The foregrounds are mostly smooth in frequency (except in the presence of effects such as polarization leakage), while the cosmological signal is not smooth in frequency, since it traces the structure of matter in the Universe. This allows component separation techniques to separate the foregrounds from the desired underlying cosmological signal, and remove them. However, these techniques can confuse the signal with the foreground, usually in the largest scale limits of the particular box, where the signal looks smooth. This leads to a loss of signal, which we try to model using a damping function across radial and transverse modes.

In this work we introduce a two-parameter damping prescription to model the effects of foreground removal. The two parameters, $N_\perp$ and $N_\parallel$, vary the scale of modes being damped by foreground removal. If $N_\perp$ and $N_\parallel$ equal zero, this corresponds to no damping. We expect the combined damping factors $N_\parallel k^{\rm min}_\parallel$ to be greater than $N_\perp k^{\rm min}_\perp$ since foreground removal mainly removes signal along the radial (LoS) direction. 

Although we could have quoted the $N_\parallel k^{\rm min}_\parallel$ and $N_\perp k^{\rm min}_\perp$ values together as just two parameters, e.g. $k^{\rm FG}_\parallel$ and $k^{\rm FG}_\perp$, we choose instead to keep this form where we have the $N_\perp$ and $N_\parallel$ parameters present. The main reason for this is that $N_\perp$ and $N_\parallel$ are independent of the box dimensions (universal for similar conditions), while $N_\parallel k^{\rm min}_\parallel$ and $N_\perp k^{\rm min}_\perp$ depend on $k^{\rm min}_\parallel$ and $k^{\rm min}_\perp$ (specific to our case, will vary depending on the particular geometry of different simulations or surveys). We expect a user to retrieve similar $N_\perp$ and $N_\parallel$ to us if using a similar foreground removal method (e.g. \fastica\ with $N_{\rm IC}=4$, \citet{Hyvrinen1999FastAR}) regardless of the box dimensions, and thus find this the most relevant parameter to quote. 

However, we note that we would expect $N_\perp$ and $N_\parallel$ to be larger for more aggressive foreground removal methods (with higher $N_{\rm IC}$), which are employed in real data to deal with more complicated foregrounds, noise and systematics (see e.g. \citet{Wolz_2016}). In particular, real foregrounds might experience polarization leakage, an effect which hinders their spectral smoothness, and would require a more aggressive $N_{\rm IC}$ choice to fully remove (see e.g. \citet{Moore_2013} for more insight on the effect of polarized foregrounds). In addition, the model is only valid for cases where the survey geometry is constant (i.e., $L_x, L_y, L_z$ are not changing with redshift). Further consideration would be needed in order to successfully apply this model to a lightcone with realistic survey geometry.

The damping term for modelling the effects of foreground removal is given in Fourier space by:
\begin{equation}\label{FGdamping}
\begin{split}
    \widetilde{B}_{\rm FG}(k, \mu) = \ & \left( 1 - \mathrm{exp}\left\{ -\left( \frac{k}{N_\perp k^{\rm min}_\perp} \right)^2 \left( 1 - \mu^2 \right) \right\} \right) \\
    \ & \times \left( 1 - \mathrm{exp}\left\{ -\left( \frac{k}{N_\parallel k^{\rm min}_\parallel} \right)^2 \mu^2 \right\} \right) \, ,
\end{split}
\end{equation}
and the power spectrum model in the presence of foreground removal effects is:
\begin{equation}\label{FullPkFG}
\begin{split}
	P_\hinospace(k^{\rm f}, \mu^{\rm f}) = \ & \frac{\widetilde{B}_\perp^2(k,\mu) \widetilde{B}_{\rm FG} (k,\mu)}{\alpha_\parallel \, \alpha_\perp^{2}} \\ 
	\ & \times \left[ \frac{\left( \overline{T}_\hinospace b_\hinospace \sigma_8 + \overline{T}_\hinospace f \sigma_8 \mu^{2}\right)^{2} P_\text{M,8}(k)}{1+\left(k \mu \sigma_\text{v} / H_{0}\right)^{2}} + P_\text{SN} \right]\, .
\end{split}
\end{equation}
When applying the multipole expansion formalism to this model, we obtain:
\begin{equation}\label{FullModMultFG}
\begin{split}
	P_{\ell}(k^{\rm f})= \ & \frac{2 \ell+1}{2} (\alpha_\parallel^{-1}\, \alpha_\perp^{-2}) \int_{-1}^{1} \diff \mu^{\rm f}\,\mathcal{L}_{\ell}(\mu^{\rm f}) \widetilde{B}_\perp^2(k,\mu) \widetilde{B}_{\rm FG}(k,\mu) \\ 
	\ & \times \left[ \frac{\left( \overline{T}_\hinospace b_\hinospace \sigma_8 + \overline{T}_\hinospace f \sigma_8 \mu^{2}\right)^{2} P_\text{M,8}(k)}{1+\left(k \mu \sigma_\text{v} / H_{0}\right)^{2}} + P_\text{SN} \right] \, .
\end{split}
\end{equation}
\section{Simulations}\label{sec:sim}
\label{sec:maths}

The simulated data we use in this investigation are the same as in C20 and we refer the reader there for a more in depth introduction. For completeness, we provide a summary here of the cosmological \hi signal simulations (\secref{sec:CosmoSig}) and the foreground simulations (\secref{sec:ForegroundSims}).

\subsection{Cosmological signal}\label{sec:CosmoSig}

The source of our simulated cosmological data is from the \textsc{MultiDark-Galaxies} data \citep{Knebe:2017eei} and the catalogue produced from the \textsc{SAGE} \citep{Croton:2016etl} semi-analytical model application. These galaxies were produced from the dark matter cosmological simulation \textsc{MultiDark-Planck} (MDPL2) \citep{Klypin:2014kpa}, which follows the evolution of 3840$^3$ particles in a cubical volume of $1 \,\text{Gpc}^3\,h^{-1}$ with mass resolution of 1.51$\,\times\, 10^9h^{-1}$M$_\odot$ per dark matter particle. The cosmology adopted for this simulation is based on \textsc{Planck}15 cosmological parameters \citep{Ade:2015xua}, with $\{\Omega_\text{M},\Omega_\text{b}, \Omega_\Lambda, \sigma_8, n_\text{s}, h\} = \{0.307, 0.048, 0.693, 0.823, 0.96, 0.678$\}.

As in our previous work (C20), we use the data from the $z=0.82$ redshift snapshot. At this redshift the box size with comoving distance dimensions $L_\text{x} = L_\text{y} = L_\text{z} = 1000\, \text{Mpc}\,h^{-1}$ approximately corresponds to a sky area of $29\times29\,\text{deg}^2$ with a redshift depth of $\Delta z = 0.5$. Using Nearest Grid Point (NGP) assignment, we bin the catalogue of galaxies onto a grid with voxel dimensions $N_\text{x}=N_\text{y}=N_\text{z}=225$. We checked that using a higher resolution grid with $N_\text{x}=N_\text{y}=N_\text{z}=512$ made no discernible difference in our analysis. From the survey volume, we have that $k_{\rm min} = 2\pi/V^{1/3} = 0.006\, h\,\text{Mpc}^{-1}$, and use bins of width $\Delta k = 0.013\, h\,\text{Mpc}^{-1}$ to avoid correlations between bins.

The SAGE catalogue we use has cold gas mass outputs for each galaxy from which we can compute a \hi brightness temperature in each pixel of our map. However, since the simulation has a finite mass resolution, the lowest mass halos ($\lesssim 10^{10}\,h^{-1}$M$_\odot$) which also contain \hi will not be properly sampled (see C20 for further discussion). This is an important limitation to consider, since it affects the observed brightness, bias, and probability distribution of \hi in our simulation. But the lowest mass haloes, albeit more abundant, contribute the least to the total brightness, and their exclusion has little effect on the bias \citep{Villaescusa_Navarro_2018,Spinelli:2019smg}. To ensure a realistic global \hi signal is present in the data we rescale the mean \hi temperature such that the \hi abundance is consistent with a value obtained in real data analyses at this redshift,  $\Omega_\hinospace \sim 4.3 \times 10^{-4}$ \citep{Masui:2012zc}. This provides our simulated data with a realistically distributed \hi signal with a mean value of $\overline{T}_\hinospace = 0.13$ mK.

We aim to emulate an upcoming SKA1-MID-like experiment \citep{Bacon:2018dui} and we therefore include simulated instrumental effects from the radio telescope beam. Using the diameter of an SKA dish ($D_\text{max} = 15$ m) we can calculate the beam size of such an experiment from
\begin{equation}\label{beamequation}
	\theta_\text{FWHM} = \frac{1.22 \, \lambda_{21} }{ D_\text{max}} (1+z) \, ,
\end{equation}
which for our case yields a beam of $\theta_\text{FWHM}=1.78\,$deg, equivalent to $R_\text{beam} = \sigma_\theta r(z) = 26.16\, \text{Mpc}\,h^{-1}$ (see \secref{sec:beammodel}). After combining our simulated \hi temperature fluctuation field and foregrounds, we convolve the final simulation cube with the telescope beam described above.

Using the \multidark\ simulation without RSD or systematic effects, we measure the linear bias of the simulation to be  $b_\hinospace (k) = \sqrt{P_\hinospace (k) / P_\text{M,linear}(k)}$, and by averaging this quantity over the large, linear $k$-scales we obtain $b_\hinospace = 1.16 \pm 0.04$. We roughly measure the upper limit on the shot noise to be $P_{\rm SN} = 2.5\,\, \text{mK}^2 \, \text{Mpc}^{3}\, h^{-3}$. This is done by looking at our measurement of $P_\hinospace (k)$ in a high resolution grid simulation of $N_\text{side}=512$, and seeing where the power spectrum `plateaus' at high $k$. At these scales the shot noise should be dominating over the faint cosmological signal, and this gives us an idea of what the upper limit of the shot noise is.

\subsection{Simulating the effect of foregrounds}\label{sec:ForegroundSims}

In order to generate realistic foregrounds we utilize the Global Sky Model (GSM) \citep{deOliveiraCosta:2008pb, Zheng:2016lul}, which extrapolates maps from real data at the desired frequency. For the redshift depth of our simulated data we can assume a frequency range of $673<\nu<903\,\text{MHz}$ and we therefore generate $N_\text{z}=225$ maps spanning this range. Therefore, the simulated foregrounds have a realistic evolving spectral index which is still smooth relative to the cosmological signal and will allow for successful component separation in the foreground clean. 

In reality, foreground signals are likely to be more complex and include contributions from free-free emission, extra-galactic point-sources and suffer effects from polarization leakage. This often requires a more aggressive foreground clean than what is typically required on a simulation from the GSM alone. In order to add additional complexity to the simulated foregrounds, we also generate realizations of diffuse emission from a model power spectrum which aims to describe different foreground sources. We follow the details outlined in Table 1 of C20 to produce these contributions, which include models of extragalactic point sources and free-free emission. We then combine these realizations with the GSM outputs to complete the full sky foreground data.

We then need to transform these full-sky maps into flat-sky data with the same dimensions as our cosmological \hi simulation. To do this we define an angular coordinate for each pixel on the flat-sky map, which we match to a pixel in the \texttt{HEALPix}\footnote{\href{http://healpix.sourceforge.net}{https://healpix.sourceforge.io/}} \citep{Gorski2005,Zonca2019} map with the closest angular coordinate. While this approach is an approximation and may affect some angular coherence in the foreground maps, it will have no impact on the foreground as a contaminant to our data. We add these flattened foreground maps onto the \hi cosmological maps to contaminate them and create the requirement for a foreground clean. 

For the foreground cleaning, we use the blind foreground removal method Fast Independent Component Analysis (\fastica) \citep{Hyvrinen1999FastAR}, and refer the reader to \citet{Wolz_2014,Cunnington:2019lvb} for a more detailed description. As discussed when introducing the foreground modelling in \secref{sec:FGmodel} and in C20, this method removes the foregrounds by assuming that the raw, uncleaned data can be written as a linear equation, where the signal can be broken up into statistically independent components:
\begin{equation}\label{ICAequation2}
	\textbf{x} = \textbf{A}\textbf{s} + \varepsilon = \sum_{i=1} ^{N_\text{IC}=m} \textbf{a}_i s_i + \varepsilon \, ,
\end{equation}
where $m$ describes the number of independent components, $\textbf{x}$ is the raw, uncleaned data, $\textbf{A}$ is the mixing matrix which describes the amplitude of the independent components, $\textbf{s}$ are the $m$ independent components, and $\varepsilon$ is the residual which includes noise and the cosmological signal. As an input, we choose $N_\text{IC} = 4$ in accordance with previous studies \citep{Chapman:2012yj,Wolz_2014,Cunnington:2019lvb,Cunnington:2020mnn}. The independent components in this case are the foregrounds, and by appropriately identifying and removing these we are left with $\varepsilon$, which contains our cosmological \hi signal.

This process of foreground removal is imperfect, and tends to confuse the signal with the foreground at large scales, especially in small $k_\parallel$ modes where the cosmological signal also appears smooth in frequency. This leads to cosmological signal being removed, which affects the amplitude of the power spectrum (see e.g. \citet{Alonso_2014} for further discussion). We show in our analysis that it is possible to account for this effect using a model with free parameters that we let vary.

\subsection{Instrumental Noise}

\begin{figure}
	\includegraphics[width=\columnwidth]{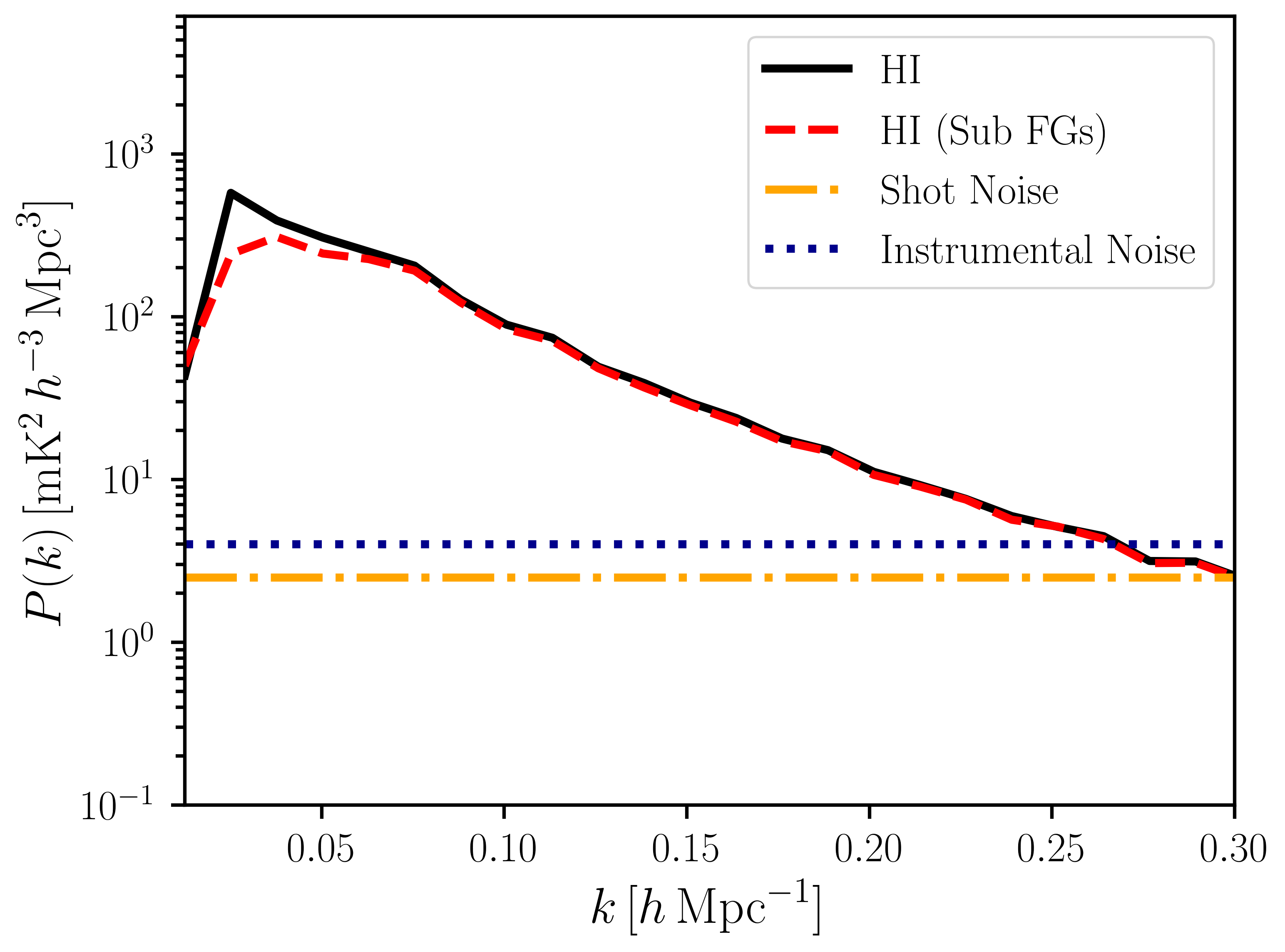}
    \caption{$\hinospace$ cosmological signal for the \multidark\ simulation at $z=0.82$ with SKA1-MID-like specifications (black solid line), foreground cleaned $\hinospace$ cosmological signal (red dashed line), the assumed instrumental noise power spectrum for such an experiment (blue dotted line), and the estimated shot noise of the simulation (orange dash-dotted line)}
    \label{fig:noisevssignal}
\end{figure}

Instrumental noise is determined by the telescope configuration. For an SKA-like single-dish experiment, we assume the pixel noise is well represented by a Gaussian random field with spread given by:
\begin{equation}\label{NoiseEq}
    \sigma_\text{pix} = T_\text{sys}\, \sqrt{\frac{4\pi\, f_\text{sky}}{\Omega_\text{beam}\,N_\text{dish}\,t_\text{obs}\,\deltadiff\nu}} \, ,
\end{equation}
from which the noise power spectrum is then given by $P_\text{N} = \sigma_\text{pix}^2 V_\text{pix}$, where $V_\text{pix}$ is the voxel volume given by:
\begin{equation}
    V_\text{pix} = \Omega_\text{beam} \int^{z + \Delta z / 2}_{z - \Delta z / 2} dz \frac{dV}{dzd\Omega} \, , 
\end{equation}
where $\Omega_{\rm beam}$ is defined on \autoref{tab:skanoise}, and
\begin{equation}
    \frac{dV}{dzd\Omega} = \frac{cr(z)^2}{H(z)}\, .
\end{equation}
We assume SKA1-MID-like parameters for the noise (see \citet{Bacon:2018dui}), shown in \autoref{tab:skanoise}. The calculated noise power spectrum from these specifications is $P_\text{N} = 4 \,\, \text{mK}^2 \text{Mpc}^3 h^{-3}$. 
\begin{table}
\centering
\begin{tabular}{| l | c | c |} 
 \hline
 Parameter & Description & Value \\
 \hline\hline
 $N_{\rm dish}$ & Number of dishes & 133 \\ 
 $D_{\rm dish}$ (m) & Dish diameter & 15 \\ 
 $t_{\rm obs}$ (hr) & Total observing time & 20,000 \\
 $\theta_{\rm FWHM}$ (deg) & Beam FWHM & 1.78 \\
 $\Omega_{\rm beam}$ (rad) & Beam solid angle $1.33\theta_{\rm FWHM}^2$ & 0.001 \\
 $f_{\rm sky}$ & Sky area coverage & 0.3 \\
 $T_{\rm sys}$ (K) & System temperature & 25 \\
 $z_{\rm eff}$ & Effective (central) redshift & 0.82 \\
 $\Delta z$ & Redshift bin width & 0.5 \\
 $\delta \nu$ (MHz) & Frequency resolution & 1 \\
 \hline
 $P_\text{N} \, (\text{mK}^2 \text{Mpc}^3 h^{-3})$ & Noise power spectrum & 4 \\ 
 \hline
\end{tabular}
\caption{Specifications for an SKA1-MID-like experiment, following \citet{Bacon:2018dui}.}
\label{tab:skanoise}
\end{table}
We plot a comparison between our \multidark\ $z=0.82$ cosmological signal in the absence of foregrounds, the foreground cleaned signal, the estimated shot noise and the instrumental noise, as seen in \autoref{fig:noisevssignal}. We see that the instrumental noise dominates over the signal for $k > 0.25\,h\,\text{Mpc}^{-1}$. We also note that pathfinder surveys for the SKA will have higher noise levels, but in this work we focus on the prospects of using \hi IM for \textit{precision} cosmology, hence why we choose to use SKA1-MID-like specifications.

\begin{figure*}
	\centering
	\includegraphics[width=2\columnwidth]{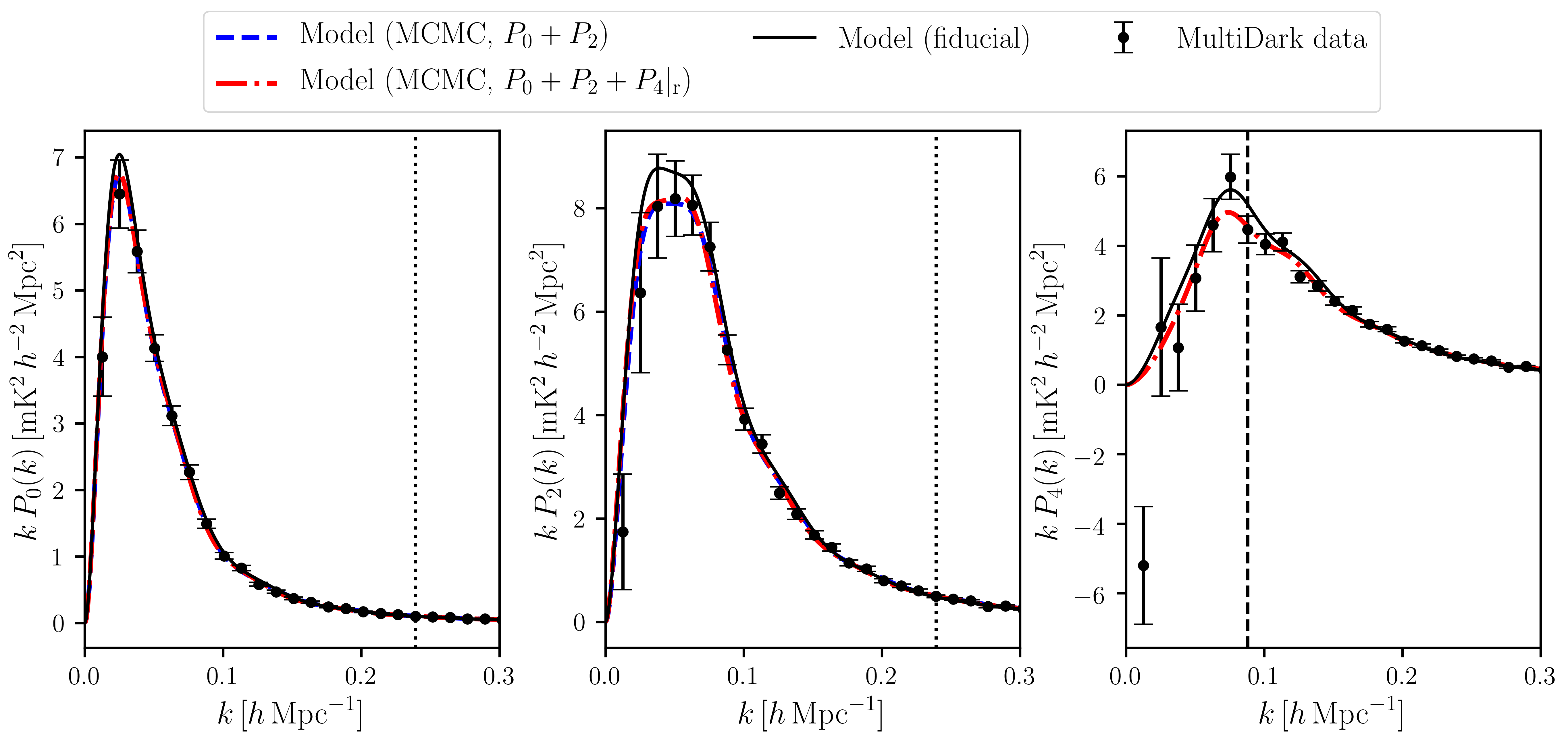}
    \caption{Model (\autoref{FullModMult}, black solid line) plotted against the $\hinospace$ power spectrum multipoles calculated from our $z=0.82$ \multidark\ \hi IM simulation (black circles), with error bars calculated using \autoref{eq:powerspecerror}. The vertical dotted line represents the limit $k_{\rm max} = 0.24 \,h\,\text{Mpc}^{-1}$ for the monopole and quadrupole, while the vertical dashed line represents the restricted $k_{\rm max} = 0.09 \,h\,\text{Mpc}^{-1}$ limit for the hexadecapole. The blue dashed line shows the best fit model obtained with the MCMC using the monopole and quadrupole, and the red dash-dotted line shows the best fit model when also adding the restricted hexadecapole.}
    \label{fig:z08results}
\end{figure*}

\begin{figure*}
	\centering
	\includegraphics[width=2\columnwidth]{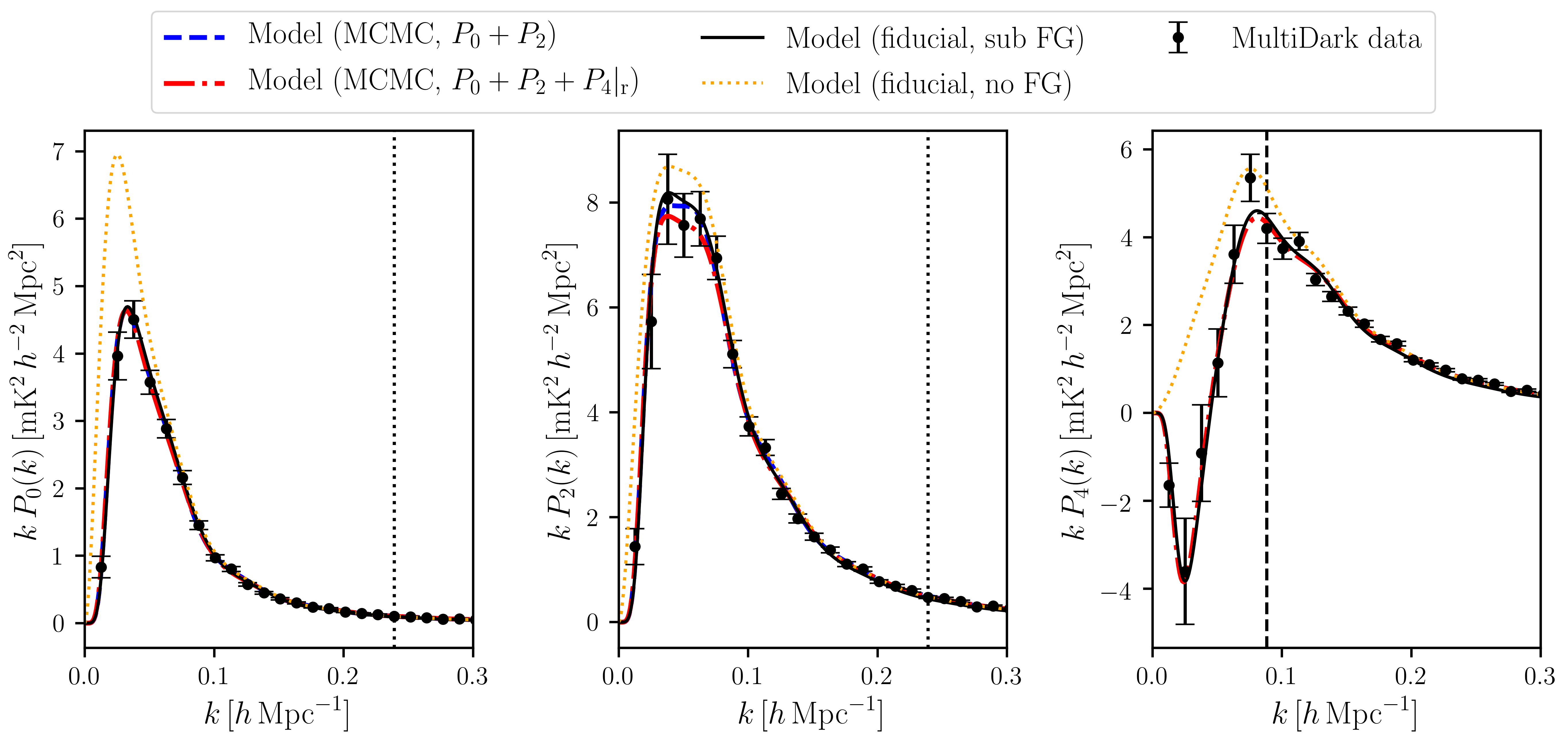}
    \caption{Foreground model (\autoref{FullModMultFG} ($N_\perp=2$, $N_\parallel=2$), black solid line) plotted against the $\hinospace$ power spectrum multipoles calculated from our simulation (black circles) in the foreground subtracted case, using $N_\text{IC} = 4$, with error bars calculated using \autoref{eq:powerspecerror}. We plot the foreground-free model (\autoref{FullModMult}, orange dotted line) for comparison. The vertical dotted line represents the limit $k_{\rm max} = 0.24 \,h\,\text{Mpc}^{-1}$ for the monopole and quadrupole, while the vertical dashed line represents the restricted $k_{\rm max} = 0.09 \,h\,\text{Mpc}^{-1}$ limit for the hexadecapole. The blue dashed line shows the best fit model obtained with the MCMC using the monopole and quadrupole, and the red dash-dotted line shows the best fit model when also adding the restricted hexadecapole.}
    \label{fig:z08resultsFG}
\end{figure*}

Combining our model, the \multidark\ simulation power spectrum results, and the noise specifications, we can see how well our model agrees with the \multidark\ data in \autoref{fig:z08results}. We also plot our foreground cleaned data ($N_\text{IC} = 4$) against the foreground model in \autoref{fig:z08resultsFG}, using guesses for the parameters $N_\perp$ and $N_\parallel$ found by eye ($N_\perp=2$, $N_\parallel=2$). For the fiducial model, we choose to use the estimated $b_\hinospace = 1.16$ and $P_{\rm SN} = 2.5\,\, \text{mK}^2 \text{Mpc}^3 h^{-3}$, and we guess by eye the velocity dispersion parameter to be $\sigma_{v} = 200\,\, \text{km/s}$. Note that when performing an MCMC analysis and checking for biased parameter results, we will not try to recover the `fiducial' values of the shot noise or velocity dispersion (since these are only rough estimates), but we will try to recover the estimated \hi bias ($b_\hinospace$) as discussed at the end of \secref{sec:CosmoSig}, which is degenerate with $\overline{T}_\hinospace$ and $\sigma_8$. The full set of fiducial cosmological parameters is outlined on \autoref{fidpars}. These are used in the fiducial model and also in the covariance matrix calculations.
\begin{table}
\centering
\begin{tabular}{| l | c |} 
 \hline
 Parameter & Fiducial value \\ 
 \hline\hline
 $\alpha_\parallel$ & 1 \\ 
 $\alpha_\perp$ & 1 \\
 $\overline{T}_\hinospace f\sigma_8$ & 0.09 $\text{mK}\,\text{Mpc}^3 h^{-3}$ \\
 $\overline{T}_\hinospace b_\hinospace \sigma_8$ & 0.12 $\text{mK}\,\text{Mpc}^3 h^{-3}$  \\
 $P_{\rm SN}$ & 2.5 $\text{mK}^2 \text{Mpc}^3 h^{-3}$ \\
 $\sigma_{v}$ & 200 $\text{km/s}$ \\
 $N_\perp$ & 2 \\
 $N_\parallel$ & 2 \\
 \hline
\end{tabular}
\caption{Fiducial model parameter values for the \hi intensity mapping \multidark\ simulation at $z=0.82$.}
\label{fidpars}
\end{table}
\subsection{Covariance matrix}

There are three main sources of error arising from the considered $\hinospace$ IM experiment: sample variance, instrumental noise and shot noise. The covariance per $k$ and $\mu$ bin (neglecting mode coupling), is:
\begin{equation}\label{eq:errors}
    \sigma^{2} (k, \mu) = \frac{\left( P_\hinospace(k, \mu) + P_N \right)^2}{N_{\rm modes}(k, \mu)}\, ,
\end{equation}
where $N_{\rm modes}(k, \mu)$ is the number of modes in each $k$ and $\mu$ bin with widths $\Delta k$ and $\Delta \mu$, respectively:
\begin{equation}
    N_{\rm modes}(k, \mu) = \frac{k^2 \Delta k \Delta \mu}{8\pi^2}V_{\rm sur}\, ,
\end{equation}
where $V_{\rm sur}$ is the volume of the survey. For a survey scanning a sky area of $\Omega_\text{tot}$ this is given by:
\begin{equation}
    V_\text{sur} = \Omega_\text{tot} \int^{z_{\rm max}}_{z_{\rm min}} dz \frac{dV}{dzd\Omega} \, .
\end{equation}
By neglecting mode coupling, we are assuming that the different $k$-bins are uncorrelated. We have assessed this assumption by comparing the multipole power spectrum errors obtained using a jackknife test to those obtained theoretically, assuming a diagonal covariance matrix, and found them to be consistent, meaning the diagonal covariance matrix is sufficient for our purposes (this assumption was also tested in C20 in the same way, and also found to be sufficient). We also note that we do not have a window function introducing correlations between different modes.

The covariance matrix of the power spectrum multipoles is comprised of the sub-covariance matrices of each multipole, and those between different multipoles (i.e. the matrix is not diagonal, as it is essential to model the non-zero covariance between multipoles, see \secref{CovSection}). The sub-covariance matrix for \hi power spectrum multipoles $\ell$ and $\ell$' is \citep{Bernal:2019lim}:
\begin{equation}\label{covLLprime2}
    C_{\ell\ell^\prime}(k) = \frac{(2\ell + 1)(2\ell^\prime + 1)}{2} \int^{1}_{-1} d\mu \, \sigma^{2} (k, \mu) \mathcal{L}_{\ell}(\mu) \mathcal{L}_{\ell^\prime}(\mu) \, .
\end{equation}
It follows from this that the total error on each multipole is given by \citep{Feldman:1993ky,Seo_2010,Battye:2012tg, Grieb:2015bia, Blake:2019ddd}:
\begin{equation}\label{eq:powerspecerror}
\begin{split}
    \sigma_{P_\ell}(k) \ & = \sqrt{\frac{(2\ell + 1)^2}{2} \int^{1}_{-1} d\mu \, \sigma^{2} (k, \mu) \mathcal{L}_{\ell}^2(\mu)} \\ \ & = (2\ell + 1)\sqrt{\int^{1}_{0} d\mu \frac{\left( P_\hinospace(k, \mu) + P_N \right)^2 \mathcal{L}_{\ell}^2(\mu)}{N_{\rm modes}(k, \mu)}} \, .
\end{split}
\end{equation}
We can compute $\left[\text{S}/\text{N}\right](k)$, the total signal-to-noise ratio per $k$ bin, as:
\begin{equation}
    \left[ \text{S}/\text{N} \right](k) ^2  = \vec{\Theta}^T (k) C^{-1}(k) \vec{\Theta} (k) \, ,
\end{equation}
where $\vec{\Theta}(k)$ is a vector describing the power spectrum per $k$ bin: 
\begin{equation}
    \vec{\Theta}(k) = [P_0 (k), P_2 (k), P_4 (k)]
\end{equation}
Similarly, the log-likelihood is proportional to the $\chi^2$ statistic:
\begin{equation}\label{logL}
\begin{split}
    \ & \log \mathcal{L} \propto -\frac{1}{2} \chi^2 \,, \\ \ & \chi^2 = \Delta\vec{\Theta}^T C^{-1}\Delta\vec{\Theta} \,,
\end{split}
\end{equation}
where $\Delta\vec{\Theta}$ is the difference between our model prediction and the measurement from our simulation for all multipoles and $k$ bins. For example, if we were considering $N_\ell = 3$ multipoles and $N_k = 20$ $k$-bins in each multipole, our covariance matrix would have dimensions $N_\ell N_k \times N_\ell N_k = 60 \times 60$ and the vector $\Delta\vec{\Theta}$ would have length $N_\ell N_k = 60$.

\begin{figure}
	\includegraphics[width=\columnwidth]{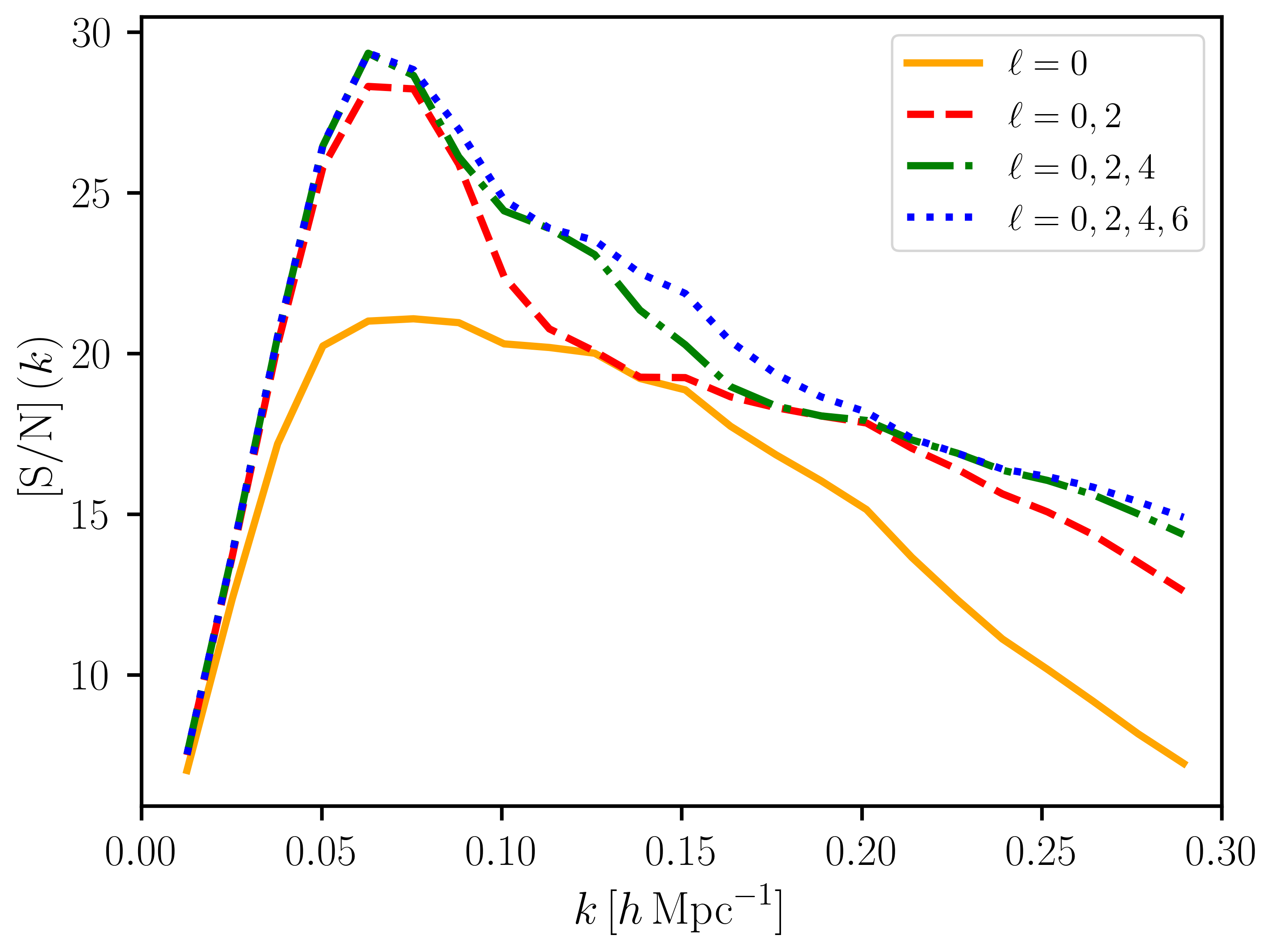}
    \caption{$\left[\text{S}/\text{N}\right]$ per $k$ bin for each combination of multipoles, for the foreground-free model (\autoref{FullModMult}).}
    \label{fig:snk024}
\end{figure}

We calculate the theoretical $\left[\text{S}/\text{N}\right](k)$ using our model (\autoref{FullModMult}) and the fiducial parameter values from the simulation (\autoref{fidpars}), as well as the assumed instrumental noise $P_\text{N} = 4 \,\, \text{mK}^2 \text{Mpc}^3 h^{-3}$. We plot the result for each combination of multipoles in \autoref{fig:snk024}. As expected, including higher order multipoles yields the best $\left[\text{S}/\text{N}\right](k)$ results, with the quadrupole adding most of the additional information at both low and high $k$ compared to the case where only the monopole is considered. We can see that the hexadecapole and 64-pole also add information, especially around $k \sim 0.15 \,h\,\text{Mpc}^{-1}$.
\section{Results}\label{sec:results}

\subsection{Model validation}

In this section, we first aim to test our model's range of validity, i.e. for which $k$ range can we trust our model to return unbiased results for the fiducial cosmological parameters of our simulation? The parameters we vary are $\{ \alpha_\parallel, \alpha_\perp, \overline{T}_\hinospace f\sigma_8,\overline{T}_\hinospace b_\hinospace \sigma_8, \sigma_v, P_{\rm SN}\}$. We know the fiducial values of the following parameters: $\{ \alpha_\parallel, \alpha_\perp, \overline{T}_\hinospace f\sigma_8,\overline{T}_\hinospace b_\hinospace \sigma_8\}$, these are outlined in \autoref{fidpars}. For all of our MCMC analyses, we keep cosmological parameters in the covariance matrix fixed to the fiducial values. We impose an upper limit of $P_{\rm SN} = 6\,\, \text{mK}^2 \, \text{Mpc}^{3}\, h^{-3}$ and $\sigma_v = 600 \, \text{km/s}$ on the shot noise and velocity dispersion parameter priors. All other priors are flat positivity priors. The MCMC analysis is performed using the publicly available \texttt{python} package \texttt{emcee}\footnote{\href{https://emcee.readthedocs.io/en/latest/}{https://emcee.readthedocs.io}} \citep{ForemanMackey:2012}. We vary our model (\autoref{FullModMult}) in the log-likelihood (\autoref{logL}) using 500 walkers and 2000 samples.

In order to validate our model, we run an MCMC analysis for different $k_{\rm max}$ limits, stopping when we see that our MCMC results are biased outside of 2$\sigma$. This test is meant to check that we are not going to $k$ values that are too large, where our theory modelling breaks down and yields biased parameter estimates. We do this only up to $k = 0.24 \,h\,\text{Mpc}^{-1}$, as beyond this we find that the signal to noise per $k$ bin drops to below 15 (see \autoref{fig:snk024}). In addition, we know that our beam starts to dominate at $k_{\rm beam} = \pi/R_{\rm beam} = 0.12 \,h\,\text{Mpc}^{-1}$, so we only consider the range up to $2k_{\rm beam} = 0.24 \,h\,\text{Mpc}^{-1}$, beyond which we assume the beam entirely dominates over the cosmological signal. We tested going beyond this $k_{\rm max}$ limit, but found no improvement on parameter uncertainties as expected.

For the case of the galaxy power spectrum, it has been shown that the monopole and quadrupole contain most of the cosmological information \citep{Taruya_2011}. The hexadecapole contains additional information, but spectroscopic galaxy surveys have found that it needs to be considered to a smaller $k_{\rm max}$ than the monopole and quadrupole due to non-linear effects (see e.g. \citet{Beutler_2016}, \citet{Markovic_2019}). Following these studies, we test whether this is also the case in \hi IM. First, we determine at which point the parameter estimation results from the MCMC become biased outside of 2$\sigma$ for the monopole and quadrupole only and find that parameter estimates are unbiased up to $k_{\rm max} = 0.24 \,h\,\text{Mpc}^{-1}$. Then, using the $k_{\rm max}$ determined for the monopole and quadrupole, we start adding the hexadecapole with different $k_{\rm max}$ limits and check when results become biased, finding that we can only include it up to a restricted range of $k_{\rm max} = 0.09 \,h\,\text{Mpc}^{-1}$.

\begin{figure}
	\centering
	\includegraphics[width=\columnwidth]{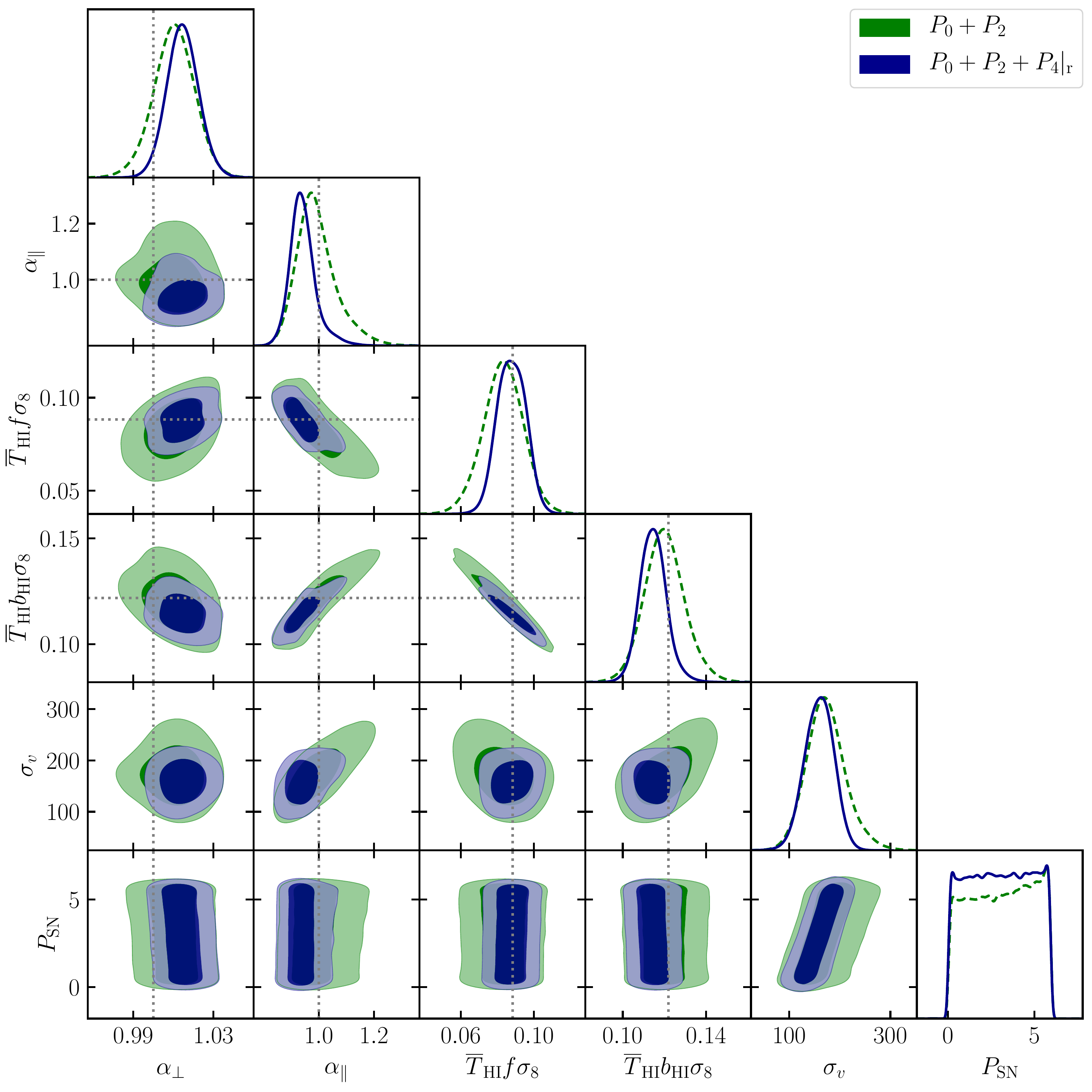}
    \caption{MCMC results for the foreground-free case, with and without the restricted hexadecaple. The dotted lines show the fiducial value for each parameter.}
    \label{fig:z08_MCMCresults}
\end{figure}

We summarise our MCMC analysis results within the model's determined range of validity in \autoref{MCMCtable_errors}, where we quote the parameter uncertainties (1$\sigma$) with and without the restricted hexadecapole. We show the MCMC results in \autoref{fig:z08_MCMCresults}. It is clear from these results that adding the hexadecapole in a restricted range decreases the error margins on the parameters. We plot our power spectrum model calculated with the best fit values from the MCMC analyses in \autoref{fig:z08results}.

Our findings are consistent with the literature. \citet{Bernal:2019lim} studied the precision of a generic line intensity mapping experiment, using a nearly identical model with the one here and synthetic data, and found that including the hexadecapole improves the precision of BAO scale measurements by 10-60\% in a Fisher matrix analysis. In the case of spectroscopic optical galaxy clustering, including the hexadecapole in an MCMC analysis has proved beneficial in decreasing uncertainties, also requiring a restricted $k$-range for the hexadecapole in order to obtain unbiased results (see e.g. \citet{Beutler_2016, Markovic_2019}). 

With regard to the BAO scale parameters, our findings are qualitatively similar to what has been found in \citet{VillaescusaNavarro:2016bao}. This study shows that while the large SKA telescope beam smears out the isotropic BAO peak signature at $z > 1$, it is possible to use the radial 21cm power spectrum to measure $H(z)$ from the BAO peak at percent level precision for redshifts $z < 2.5$. Our measurement of the radial AP effect parameter, which is a function of $H(z)$, instead relies on the overall shape of the 21cm power spectrum, which is susceptible to systematic effects. Nonetheless, we considered systematic effects due to the beam and foreground removal in our analysis, and found similar, unbiased sub-10\% percent level constraints on the radial AP parameter related to the expansion rate. In addition, even in the presence of a large beam, we also find sub-1\% percent level constraints on the transverse AP parameter related to the angular distance.

\begin{table}
	\centering
	\begin{tabular}{| l | cc | c |} 
        \multicolumn{4}{c}{Marginalised 1$\sigma$ percent errors from MCMC} \\
		\hline
		Parameter & $P_0 + P_2$ & $P_0 + P_2 + P_4|_{\rm r}$ & $P_0 + P_2$ (diag) \\
        \hline\hline
		$\alpha_\perp$ & 1.0\% & 0.8\% & 2.9\% \\
		$\alpha_\parallel$ & 7.6\% & 5.3\% & 6.1\% \\
		$\overline{T}_\hinospace f \sigma_8$ & 13.3\% & 8.8\% & 14.3\% \\
		$\overline{T}_\hinospace b_\hinospace \sigma_8$ & 8.1\% & 5.7\% & 9.7\% \\
		\hline
	\end{tabular}
    \caption{Marginalised 1$\sigma$ percent error for the parameters in our model, as found by the MCMC with and without the restricted hexadecapole in the foreground-free case. Including the hexadecapole in a restricted range improves the errors. The last column shows results when we assume the covariance matrix is diagonal.}
    \label{MCMCtable_errors}
\end{table}

\subsection{Covariance Matrix}\label{CovSection}

Here we discuss our choice of using a non-diagonal covariance matrix (\autoref{covLLprime2}), meaning we consider covariance between different multipoles. We first describe the details of this non-diagonal covariance matrix. 

The full covariance matrix is symmetric, and is composed of each diagonal sub-covariance matrix (diagonal since we neglect mode coupling, and since we do not have a survey window function introducing further correlations between modes). Each sub-covariance matrix has dimensions $N_k \times N_k$ where $N_k$ is the number of $k$ bins considered per multipole, so overall the full covariance matrix has dimensions $N_\ell N_k \times N_\ell N_k$ (where $N_\ell$ is the number of multipoles being considered). That is:
\begin{equation}
    C = 
    \begin{bmatrix}
        C_{00} & C_{02} & C_{04} \\
         & C_{22} & C_{24} \\
            &   & C_{44}
    \end{bmatrix}
    ,
\end{equation}
where each sub-covariance matrix is:
\begin{equation}\label{cov_elements}
    C_{\ell\ell^\prime} = 
    \begin{bmatrix}
        C_{\ell\ell^\prime}(k_0) & 0  & 0  \\
        0 & C_{\ell\ell^\prime}(k_1) &  0 \\
          0  &  0 & \ddots
    \end{bmatrix}.
\end{equation}
The full diagonal covariance matrix, in the case where we do not consider the covariance between different multipoles, also has dimensions $N_\ell N_k \times N_\ell N_k$ and is given by:
\begin{equation}
    C_{\rm diag} = 
    \begin{bmatrix}
        C_{00} & 0 & 0 \\
        0 & C_{22} & 0 \\
          0  & 0  & C_{44}
    \end{bmatrix}
    ,
\end{equation}
where each sub-covariance matrix is also given by \autoref{cov_elements}, but only for the case of $\ell = \ell^{\prime}$. 

For more detailed discussion of the covariance of galaxy power spectrum multipoles under the Gaussian assumption, and in particular the significance of the covariance between different multipoles, see e.g. \citet{Grieb:2015bia} and \citet{Blake:2019ddd}.

\subsubsection{Effect of the telescope beam}

\begin{figure}
	\includegraphics[width=\columnwidth]{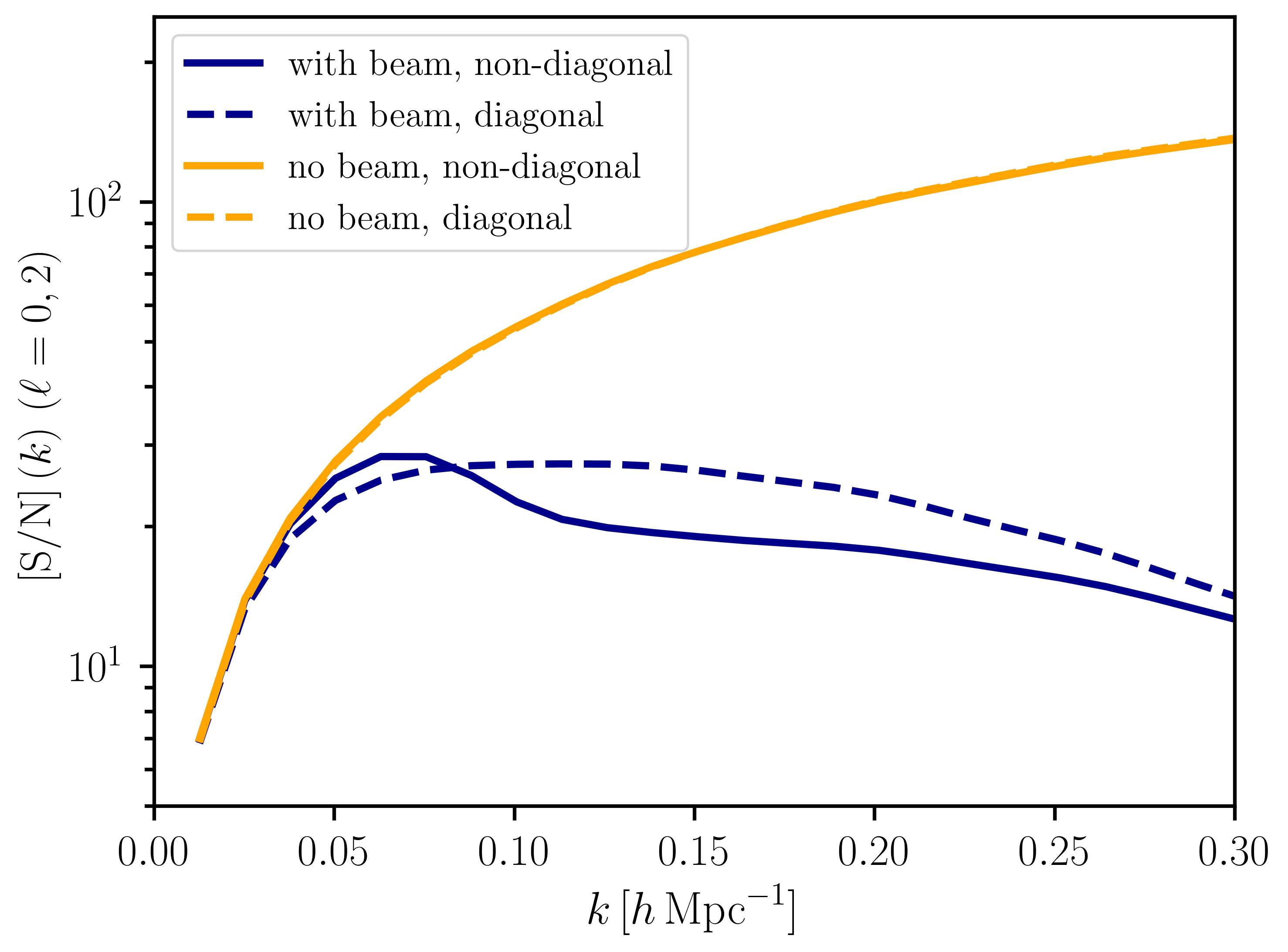}
    \caption{$\left[\text{S}/\text{N}\right]$ per $k$ bin for multipoles $\ell = 0,2$. Orange colors represent the case of no telescope beam, dark blue represents our case of a telescope beam with $R_{\rm beam} = 26.2 \,\,\text{Mpc}\,h^{-1}$. Dashed lines are the diagonal covariance matrix cases, while the solid lines include the covariance between different multipoles.}
    \label{fig:snk024diagvsnondiag}
\end{figure}

In order to compare the diagonal and non-diagonal cases, we calculate the $\left[\text{S}/\text{N}\right]$ per $k$ bin for each case with and without a telescope beam damping term. Results for $\ell = 0,2$ are given in \autoref{fig:snk024diagvsnondiag}. For the case where the telescope beam damping term is present, we can see that using a non-diagonal covariance matrix makes a difference at both low and high $k$. At low $k$, including covariance between multipoles seems to increase the $\left[\text{S}/\text{N}\right]$, while for higher $k$ it decreases it. For the case where we do not include the telescope beam, we can see that the $\left[\text{S}/\text{N}\right]$ per $k$ bin does not differ significantly between including or excluding off-diagonal terms.

\begin{figure*}
	\centering
	\includegraphics[width=2\columnwidth]{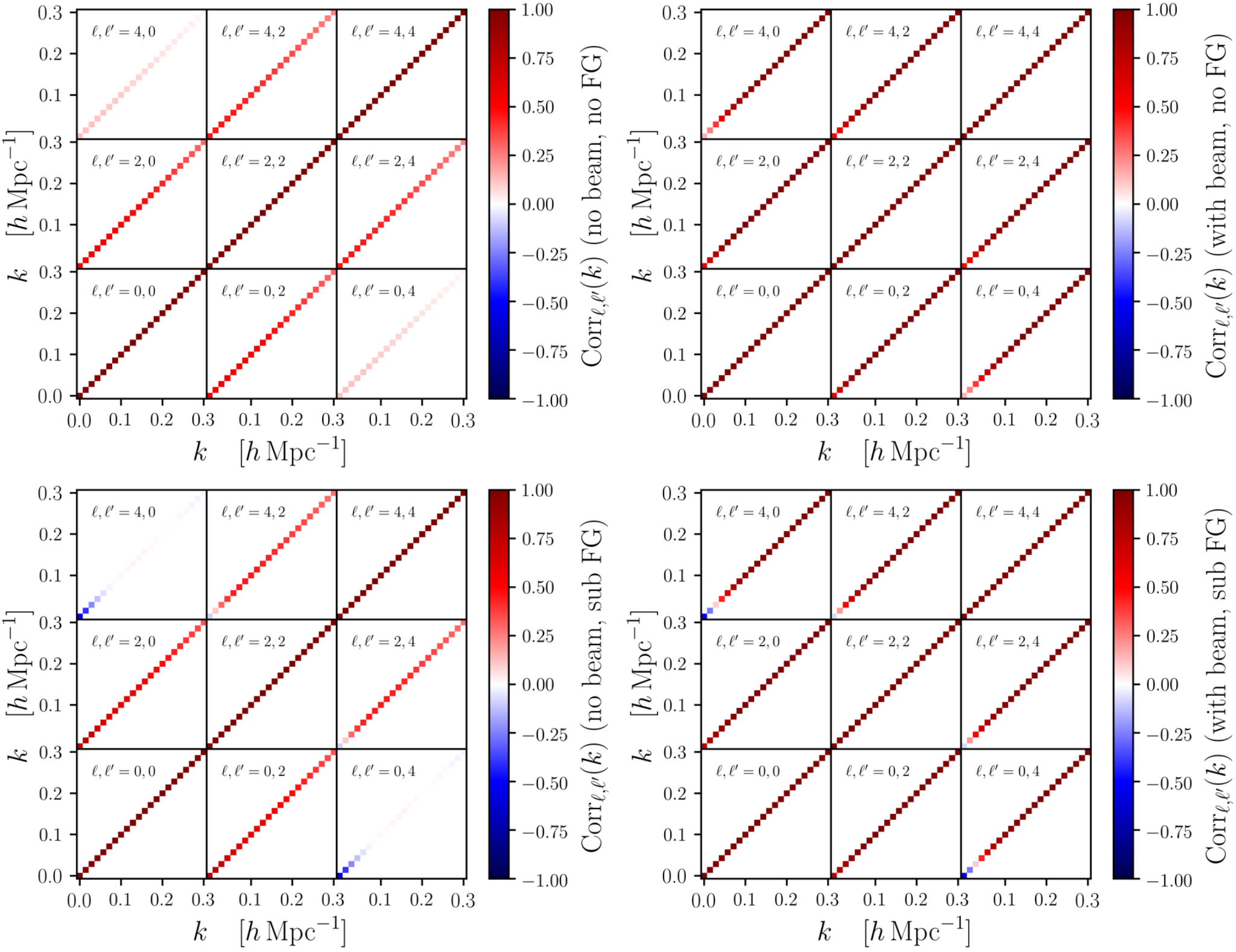}
    \caption{The correlation matrix for multipoles $\ell=0,2,4$ up to $k = 0.3\,h\,\text{Mpc}^{-1}$, excluding (\textit{left}) and including (\textit{right}) the effects of a telescope beam with $R_{\rm beam}=26.2\,\text{Mpc}\,h^{-1}$. \textit{Top}: Foreground-free. \textit{Bottom}: Including the effects of foreground removal with $N_\perp=2, N_\parallel=2$.}
    \label{fig:corrbeamvsnobeam}
\end{figure*}

It is interesting to also compare how the telescope beam changes correlations between different multipoles. We can calculate the correlation matrix from the covariance matrix as:
\begin{equation}
    \text{Corr}_{\ell\ell^\prime}(k) = \frac{C_{\ell\ell^\prime}(k)}{\sqrt{C_{\ell\ell}(k) C_{\ell^\prime \ell^\prime}(k)}}\, .
\end{equation}
We plot the correlation matrix in the case of no telescope beam and compare it to the case of including a telescope beam with $R_{\rm beam} = 26.2 \,\,\text{Mpc}\,h^{-1}$ (the same beam used in our simulation) in \autoref{fig:corrbeamvsnobeam} (top-row). We can clearly see that the presence of the telescope beam increases the correlations in the off-diagonal terms, i.e. the beam makes the different multipoles more correlated. This is because the telescope beam damping term breaks the orthogonality of the multipoles. See \appref{CovAppendix} for a more detailed discussion and derivation. 

We demonstrate that these differences due to the telescope beam in the correlation matrix and in the $\left[\text{S}/\text{N}\right]$ per $k$ carry over to an MCMC analysis in the foreground-free case. We perform the MCMC analysis for the monopole and quadrupole at the determined $k_{\rm max}=0.24\,h\,\text{Mpc}^{-1}$ for the non-diagonal and diagonal covariance matrix cases, and quote results for both cases on \autoref{MCMCtable_errors}. From these results we can determine that the errors on our model parameters, as seen on \autoref{MCMCtable_errors}, increase when ignoring the covariance between different multipoles (with the exception of the $\alpha_\parallel$ parameter, where the error slightly decreases). In both cases, we obtain unbiased parameter estimates.

In the case of considering only a diagonal covariance matrix, the percentage uncertainties on the $\alpha_\perp$ parameter (\autoref{MCMCtable_errors}) are approximately 2 times smaller than those on $\alpha_\parallel$, which is consistent with results from optical galaxy surveys (see e.g. \citet{gilmarn2020completed}). When considering the covariance between different multipoles, the fractional uncertainties on $\alpha_\perp$ become approximately 5 times smaller than on $\alpha_\parallel$. We attribute this to the telescope beam having a significant effect on the correlation between different multipoles, but leave further investigation of the effect of the telescope beam and non-diagonal covariance matrix on the AP parameters for future work.

We conclude that, when in the presence of a telescope beam, using a non-diagonal covariance matrix is important for the following reasons:
\begin{itemize}[leftmargin=*]
    \item It makes a difference in the $\left[\text{S}/\text{N}\right]$ per $k$ bin result, increasing $\left[\text{S}/\text{N}\right]$ in the low $k$ limit but decreasing it in the higher $k$ limit;
    \item The different multipoles are non-negligibly correlated due to the telescope beam;
    \item It decreases the uncertainties in most cosmological parameter estimates obtained using MCMC.
\end{itemize}

\subsection{Effect of foreground removal}

Here we aim to assess the validity of our foreground model. For our simulation, we have $k^{\rm min}_\perp = 0.004 \,h\,\text{Mpc}^{-1}$ and $k^{\rm min}_\parallel = 0.006 \,h\,\text{Mpc}^{-1}$, making the foreground damping scales $N_\perp k^{\rm min}_\perp = 0.009 \,h\,\text{Mpc}^{-1}$ and $N_\parallel k^{\rm min}_\parallel = 0.013 \,h\,\text{Mpc}^{-1}$ respectively. We note that, although we find $N_\perp=2, N_\parallel=2$ to fit our data well by eye, this does not mean that the same amount of power is being damped on both the perpendicular and parallel to the LoS directions. Indeed, when looking at the damping scales, we can see that $N_\parallel k^{\rm min}_\parallel > N_\perp k^{\rm min}_\perp$, meaning that more power is being damped in the parallel to the LoS direction.

\begin{figure*}
	\centering
	\includegraphics[width=2\columnwidth]{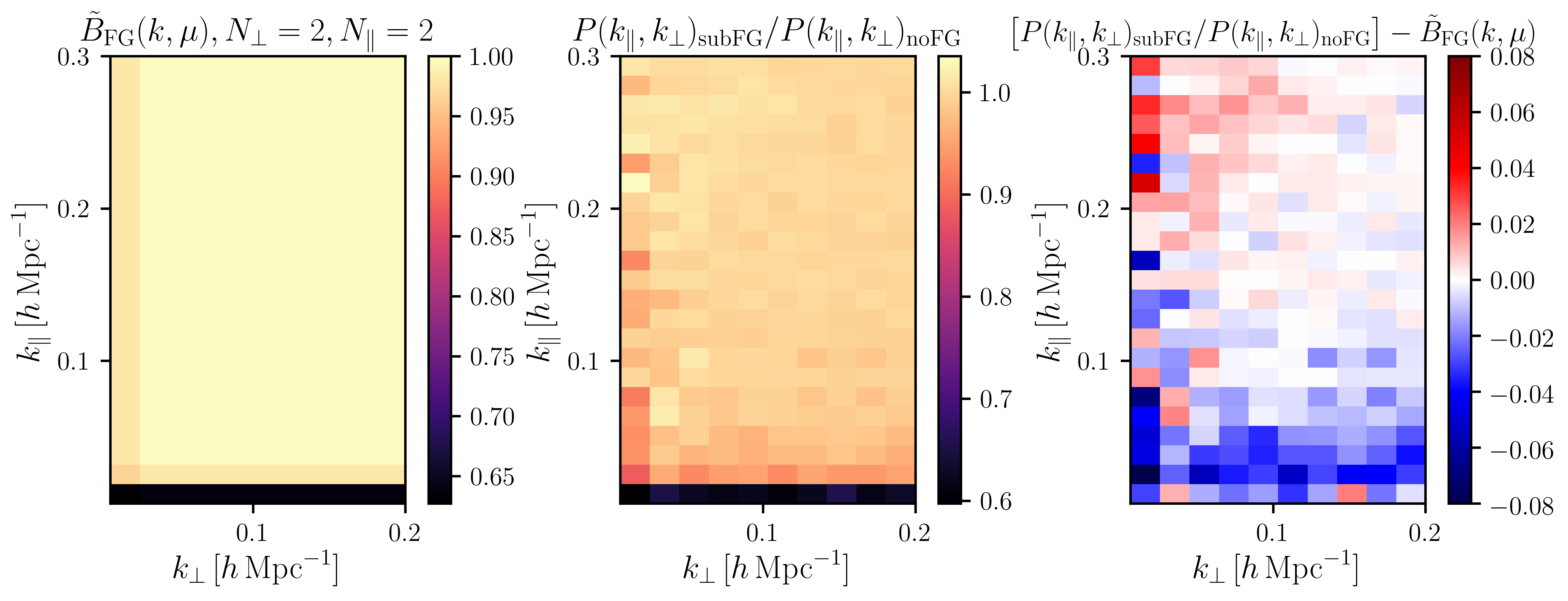}
    \caption{\textit{Left:} Foreground damping model $\widetilde{B}_\text{FG}(k, \mu)$ (\autoref{FGdamping}), with $N_\perp=2, N_\parallel=2$. \textit{Center:} Ratio of foreground removed to foreground-free $P(k_\perp,k_\parallel)$. \textit{Right:} Difference between the left and middle panels, as a proxy for how accurate our foreground damping model describes our data.}
    \label{fig:2DpkFG}
\end{figure*}

To motivate our foreground model further, we attempt to compare it to a measurement of the power spectrum decomposed into perpendicular and parallel modes, $P(k_\perp,k_\parallel)$. We compare $P(k_\perp,k_\parallel)$ in the foreground-free case to $P(k_\perp,k_\parallel)$ in the foreground removed case by plotting the ratio of these, and compare it to our foreground model $\widetilde{B}_\text{FG}(k, \mu)$ (\autoref{FGdamping}) with $N_\perp=2, N_\parallel=2$ (\autoref{fig:2DpkFG}). We also plot the difference between these, finding that they are in agreement and that differences are below 10\% on all scales. As seen in \autoref{fig:2DpkFG}, both our model and the data show more power being damped on small $k_\parallel$ modes, as expected. This comparison was also carried out in \citet{Cunnington:2020wdu}, which found similar agreement with a similar model.

We perform an MCMC analysis with the foreground subtracted data in four different cases, first only considering the monopole and quadrupole only up to $k_{\rm max} = 0.24 \,h\,\text{Mpc}^{-1}$ and later considering the inclusion of the hexadecapole up to the limit found in the foreground-free case $k_{\rm max} = 0.09 \,h\,\text{Mpc}^{-1}$. We check that up to these limits, results are unbiased as in the foreground-free case (except for case 1, where results are biased). For most cases, we are varying the parameters $\{ \alpha_\parallel, \alpha_\perp, \overline{T}_\hinospace f\sigma_8,\overline{T}_\hinospace b_\hinospace \sigma_8, \sigma_v, P_{\rm SN}\}$ and use the same priors limits as in the foreground-free MCMC analysis case. In one of the cases (case 3), we vary two additional parameters from our foreground model, namely $N_\perp, N_\parallel$, bringing the full list of parameters we vary to $\{ \alpha_\parallel, \alpha_\perp, \overline{T}_\hinospace f\sigma_8,\overline{T}_\hinospace b_\hinospace \sigma_8, \sigma_v, P_{\rm SN}, N_\perp, N_\parallel\}$. For $N_\perp, N_\parallel$, we impose flat positivity priors.

\textbf{Case 1: The foreground-free model}. First, we consider the foreground-free model (\autoref{FullModMult}) to demonstrate how it yields biased parameter estimates (specifically, the parameters $\alpha_\perp$, $\overline{T}_\hinospace f \sigma_8$, and $\overline{T}_\hinospace b_\hinospace \sigma_8$ become biased outside of the 2$\sigma$ limit). In this case we do not include the foreground model in the covariance matrix.

\textbf{Case 2: The fixed foreground model}. Next, we consider our foreground model (\autoref{FullModMultFG}) and keep the $N_\perp$, $N_\parallel$ parameters fixed to the best fit guesses found by eye ($N_\perp=2$, $N_\parallel=2$). In this case we include the foreground model in the covariance matrix.

\textbf{Case 3: The varied foreground model}. Here we consider the foreground model (\autoref{FullModMultFG}) but let $N_\perp$ and $N_\parallel$ be nuisance parameters that we vary. Here we also include the foreground model in the covariance matrix. We also compare with the case of not including the foreground model in the covariance matrix, which causes the covariance matrix values to be larger and consequently we find that this increases errors in the parameters but does not cause them to become biased. This is relevant to the case of a real data analysis, where we would not know the fiducial $N_\perp$ and $N_\parallel$ values in advance to fix in the covariance matrix, and would probably need to adopt this more conservative case. If end-to-end simulations were available that allowed for $N_\perp$ and $N_\parallel$ to be accurately determined for real data, the less conservative case could be adopted instead. Alternatively, an iterative process could also be employed with real data. We would start by assuming $N_\perp$ and $N_\parallel$ in the covariance matrix, run a parameter estimation, re-generate the covariance based on the best-fitting values, and re-run the parameter estimation until there is convergence.

\textbf{Case 4: The $k_{\rm min}$-cut model}. Finally, we investigate what happens when we exclude the largest scales where foreground subtraction has the most impact. We impose a $k_{\rm min}$ limit on the foreground subtracted data and try to recover cosmological parameters using the foreground-free model (\autoref{FullModMult}), which we have seen would yield biased parameter results if considering the full $k$-range (case 1). Here we do not include the foreground model in the covariance matrix. We find that the limit $k_{\rm min} = 0.05 \,h\,\text{Mpc}^{-1}$ is sufficient to then recover unbiased parameter estimates, and we quote the uncertainties on these on \autoref{MCMCtable_errors_FG}. Note that including the hexadecapole with this cut, or a more restricted cut, yields biased results due to the considerable impact that foreground removal has on the hexadecapole. For all parameters, the uncertainty obtained with the $k_{\rm min}$-cut method is larger than the uncertainty obtained using any other method. Furthermore, the varied $N_\perp, N_\parallel$ method yields smaller uncertainties, and does not require a prior selection of a $k_{\rm min}$ limit.

\begin{figure}
	\includegraphics[width=\columnwidth]{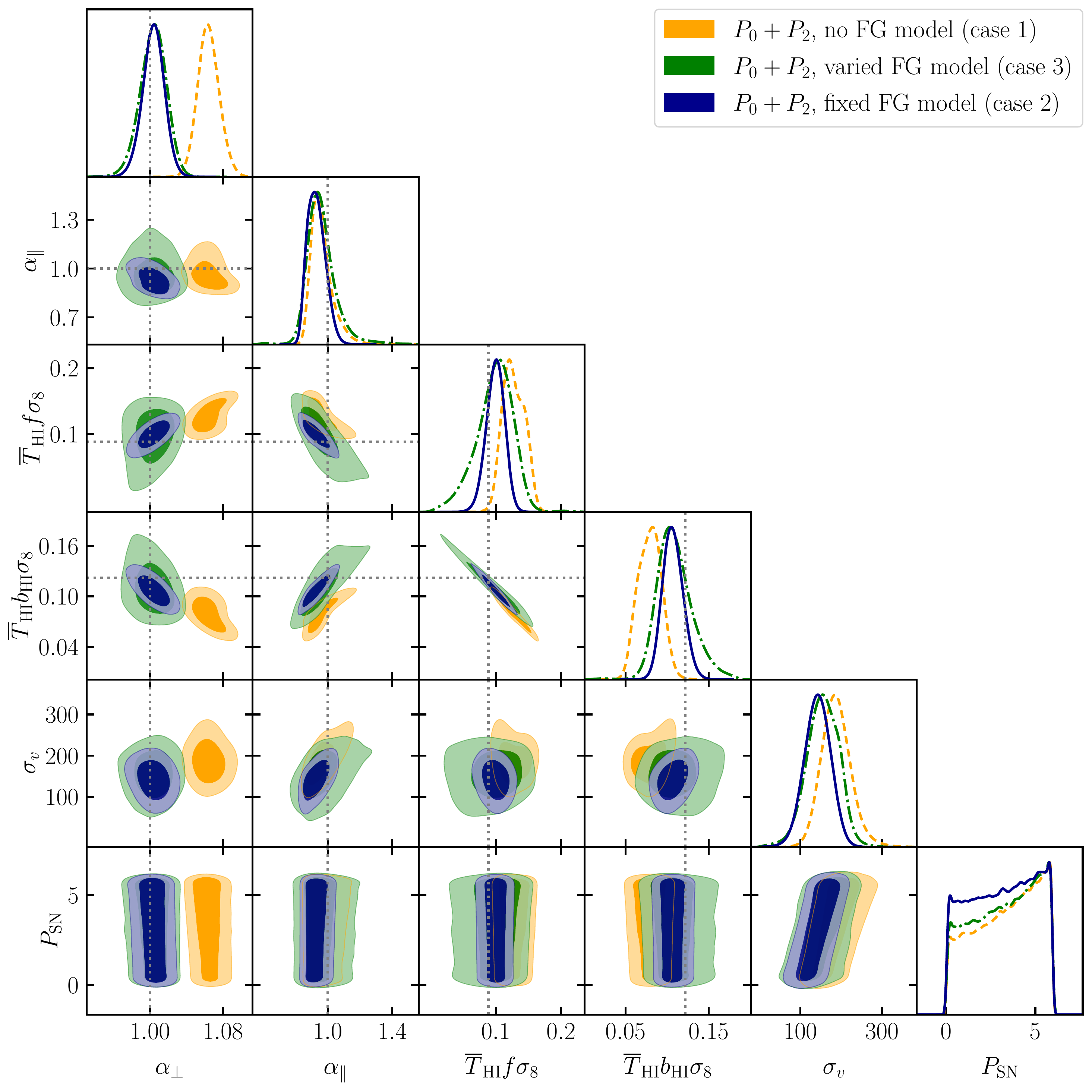}
    \caption{MCMC results for foreground subtracted data and three different model considerations (cases 1-3), using the monopole and quadrupole up to $k_{\rm max} = 0.24 \,h\,\text{Mpc}^{-1}$. The dotted lines show the fiducial value for each parameter.}
    \label{fig:MCMC_diffFGcases}
\end{figure}

Results from the MCMC analyses for cases 1 to 3 can be found in \autoref{fig:MCMC_diffFGcases}. We quote the different uncertainties on the parameters for all cases on \autoref{MCMCtable_errors_FG}.

\begin{figure*}
	\includegraphics[width=2\columnwidth]{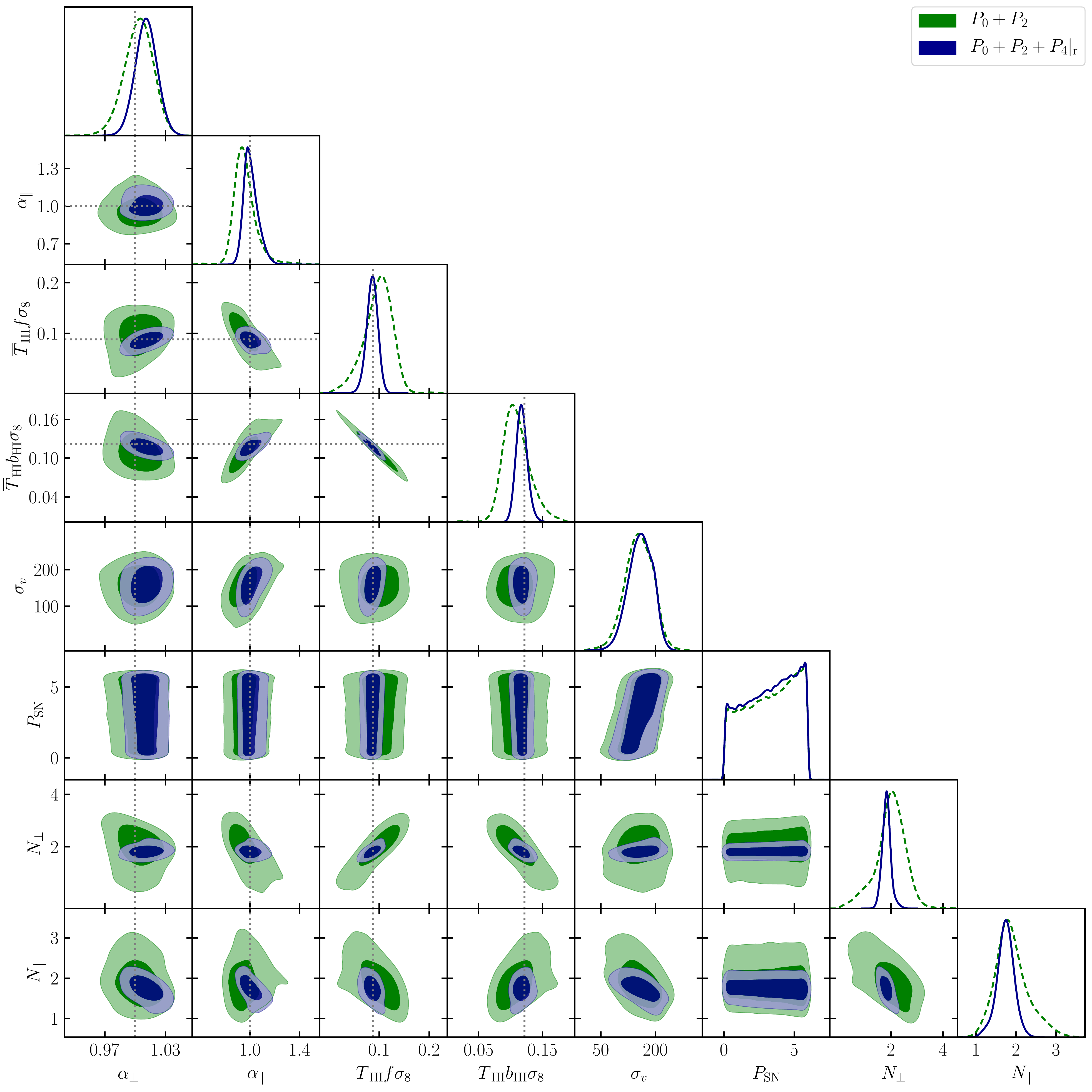}
    \caption{MCMC results for the foreground subtracted data and our varied foreground model (case 3), with and without the restricted hexadecapole. The dotted lines show the fiducial value for each parameter.}
    \label{fig:MCMC_FG_hexavsnohexa}
\end{figure*}

We now consider our varied foreground model in more depth for case 3. Still letting $N_\perp$ and $N_\parallel$ be nuisance parameters and including the foreground model in the covariance matrix, we compare the MCMC analysis results when excluding or including the hexadecapole at a restricted range of $k_{\rm max} = 0.09 \,h\,\text{Mpc}^{-1}$. Results can be found in \autoref{fig:MCMC_FG_hexavsnohexa}, and 1$\sigma$ uncertainties on \autoref{MCMCtable_errors_FG}. We can see that as in the foreground-free case, adding the hexadecapole at a restricted range still allows us to retrieve unbiased cosmological parameters with smaller uncertainties than without it. Including the hexadecapole also seems to make the posteriors more Gaussian-like, in particular for $N_\perp$ and $N_\parallel$. This motivates further the inclusion of the hexadecapole in parameter estimation analyses, particularly in foreground removed data. We plot our foreground model with best fit parameters from the case 3 MCMC analyses (with and without the hexadecapole) in \autoref{fig:z08resultsFG}.

\begin{table*}
	\centering
	\begin{tabular}{ | l | ccccc | c | } 
        \multicolumn{6}{c}{Marginalised 1$\sigma$ percent errors from MCMC, foreground subtracted case} \\
        \hline
        \multirow{ 2}{*}{Parameter} & \multicolumn{5}{c|}{$P_0 + P_2$} & \multicolumn{1}{c|}{$P_0 + P_2 + P_4|_{\rm r}$} \\
		 & No FG & Fixed $N_\perp, N_\parallel$ & Varied $N_\perp, N_\parallel$ & Varied $N_\perp, N_\parallel$ (no FG covariance) & $k_{\rm min}$-cut & Varied $N_\perp, N_\parallel$ \\
        \hline\hline
		$\alpha_\perp$ & 1.1\% & 1.2\% & 1.5\% & 2.1\% & 2.6\% & 1.1\% \\
		$\alpha_\parallel$ & 7.4\% & 5.9\% & 10.3\% & 11.0\% & 29.0\% & 5.9\% \\
		$\overline{T}_\hinospace f \sigma_8$ & 13.0\% & 14.0\% & 28.9\% & 34.1\% & 44.4\% & 13.3\% \\
		$\overline{T}_\hinospace b_\hinospace \sigma_8$ & 17.4\% & 11.5\% & 20.3\% & 21.1\% & 38.2\% & 7.8\% \\
		$N_\perp$ & N/A & N/A & 29.0\% & 43.4\% & N/A & 9.4\% \\
		$N_\parallel$ & N/A & N/A & 22.7\% & 30.7\% & N/A & 12.6\% \\
		\hline
	\end{tabular}
    \caption{Marginalised 1$\sigma$ percent error for the parameters in our model for the foreground removed case, as found by the MCMC with and without different foreground removal effects considerations.}
    \label{MCMCtable_errors_FG}
\end{table*}

\subsubsection{Covariance matrix}

We find that the effect of foreground removal significantly impacts the covariance matrix, and discuss this effect further keeping in mind that this is specific to our choice of modelling, simulations and survey specifications (which determine the instrumental noise level). For real data, one would need realistic end-to-end simulations specific to a given experiment in order to robustly include the effects of foregrounds in the covariance matrix. 

We compare the theoretical covariance matrix with and without foreground removal effects included. As seen in \autoref{fig:corrbeamvsnobeam}, where we include the foreground removal model with $N_\perp=2$, $N_\parallel=2$ in the covariance matrix, this makes a significant difference in the correlation between the different multipoles' large scale modes.

When performing the different MCMC analyses with the monopole and quadrupole for the foreground subtracted case, we showed in case 3 that including the foreground model in the covariance matrix decreases errors in the cosmological parameters of interest. However, it requires knowing the best fit $N_\perp, N_\parallel$ beforehand, which might be unlikely with real data. Nonetheless, this test shows that we are able to retrieve unbiased parameter estimates using our foreground model in \textit{either} case of including or not including foreground removal effects in the covariance matrix, but with different resulting parameter uncertainties.

As an additional indicator of how well our models fit the simulation measurements, we have also looked at the reduced $\chi^2$ of our best-fits: $\chi^2_{\rm red} = \chi^2/\text{dof}$, where dof is the degrees of freedom found by subtracting the model parameters from the number of data points. It is useful to look at $\chi^2_{\rm red}$ because if it is much larger than 1, that usually indicates an incorrect model or underestimated errors, and if it is much smaller than 1 then we could be overestimating the errors/overfitting. We calculate the $\chi^2_{\rm red}$ for the monopole, quadrupole (up to $k = 0.24 \,h\,\text{Mpc}^{-1}$) and for the hexadecapole (up to $k = 0.09 \,h\,\text{Mpc}^{-1}$). For the foreground-free case, we find $\chi^2_{\rm red} \simeq 0.6$. For the foreground removed case, we find $\chi^2_{\rm red} \simeq 1$ when including the foreground removal effects in the covariance matrix. As expected, when trying to fit the foreground-free model to foreground removed data, we obtain a best-fit $\chi^2_{\rm red} \simeq 2.6$ (and heavily biased cosmological parameter estimates, see \autoref{fig:MCMC_diffFGcases}), confirming the need for an appropriate foreground model.

Although the reduced $\chi^2$ is a useful check and indicator that our model is appropriate for fitting our measurements, our main findings and model validation come from the MCMC analyses, which recovers the cosmological parameters within 2$\sigma$ errors of the fiducial values.

\subsection{Higher order multipoles}

\begin{figure}
	\includegraphics[width=\columnwidth]{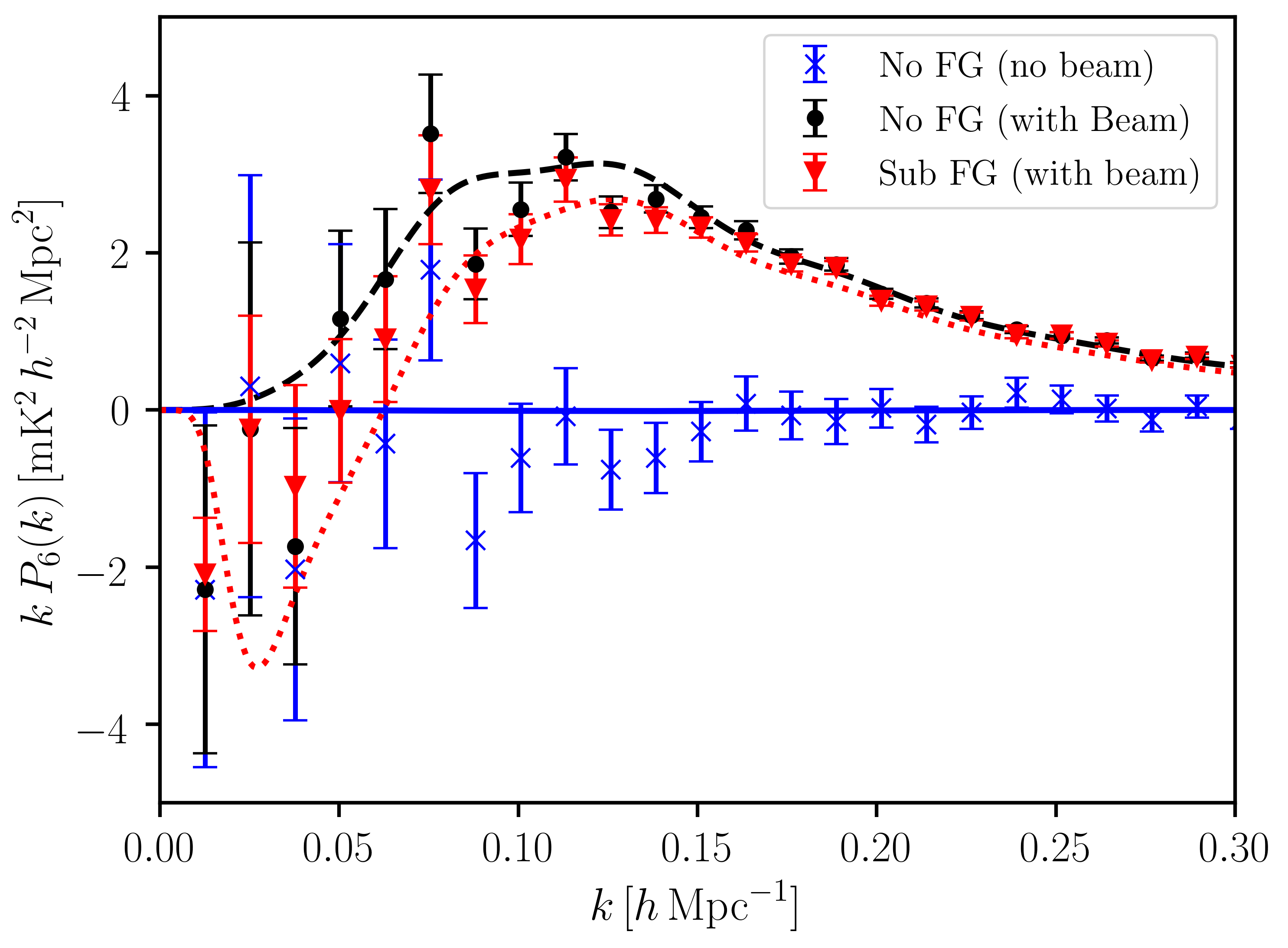}
    \caption{The 64-pole model plotted against the measurements from our simulation without any beam or foreground removal effects (blue crosses), with the telescope beam effect included (black circles) and with the telescope beam and foreground removal effects ($N_\perp=2,N_\parallel=2$) included (red triangles).}
    \label{fig:64pole_beamFG}
\end{figure}

\begin{figure}
	\includegraphics[width=\columnwidth]{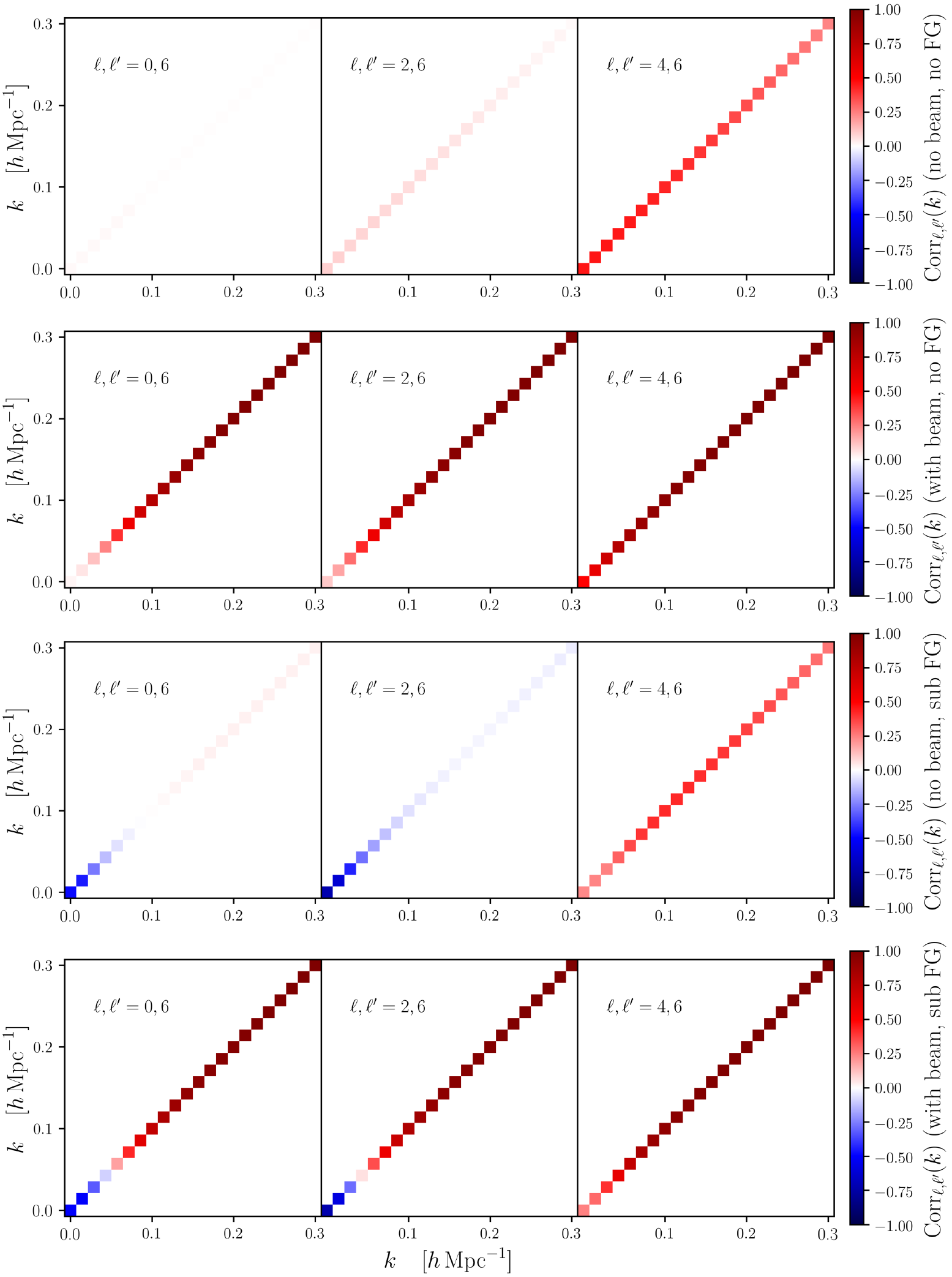}
    \caption{The 64-pole off-diagonal components of the correlation matrices, with and without telescope beam and foreground removal effects ($N_\perp=2,N_\parallel=2$).}
    \label{fig:64pole_corr}
\end{figure}

Here we investigate the effect of including higher order multipoles in our analysis (see e.g. \citet{Chuang_2013} and \citet{Uhlemann_2015} for examples of higher order multipoles being considered in galaxy and halo two-point correlation functions, respectively). The $P_{6}$ 64-pole (or hexacontatetrapole) encompasses non-linear velocity information, vanishing in the case of considering only the linear Kaiser RSD effect \citep{Kaiser:1987qv}. In the case of our simulations (ignoring beam and foreground effects), where non-linear velocity effects are present (such as the FoG effect), the 64-pole is non-zero but still expected to be very small. However, we show in \autoref{fig:64pole_beamFG} that the 64-pole is significantly affected by the telescope beam and foreground removal effects similarly to the other multipoles, meaning its signal is boosted due to these systematic and instrumental effects.
Regarding the correlation matrix, we again find that the beam and foreground removal effects significantly affect the correlations between the 64-pole and other multipoles, as seen in \autoref{fig:64pole_corr}.

We test the effect of adding the 64-pole to our parameter estimation pipeline, first in the foreground-free case. We find that we can add the 64-pole up to the same restricted range as the hexadecapole ($k_{\rm max} = 0.09 \,h\,\text{Mpc}^{-1}$), and that it does improve results by decreasing the errors on our parameters while maintaining the estimates unbiased within 2$\sigma$ (see \autoref{MCMCtable_errors_P6}).
\begin{table}
	\centering
	\begin{tabular}{| l | ccc |} 
        \multicolumn{4}{c}{Marginalised 1$\sigma$ percent errors, $P_0 + P_2 + P_4|_{\rm r}+ P_6|_{\rm r}$} \\
		\hline
		Parameter & No FG & Sub FG (FG cov) & Sub FG (no FG cov) \\
        \hline\hline
		$\alpha_\perp$ & 0.8\% & 1.2\% & 1.9\% \\
		$\alpha_\parallel$ & 4.1\% & 5.3\% & 10.6\% \\
		$\overline{T}_\hinospace f \sigma_8$ & 8.4\% & 25.1\% & 35.0\% \\
		$\overline{T}_\hinospace b_\hinospace \sigma_8$ & 4.9\% & 13.0\% & 16.6\% \\
		$N_\perp$ & N/A & 21.9 & 24.8\% \\
		$N_\parallel$ & N/A & 22.7 & 14.5\%\\
		\hline
	\end{tabular}
    \caption{Marginalised 1$\sigma$ percent error for the parameters in our model, as found by the MCMC analysis with the restricted 64-pole in the foreground-free and foreground removed cases.}
    \label{MCMCtable_errors_P6}
\end{table}

We also tested whether including the 64-pole in the foreground removed case would make a difference, and indeed it did. When we added the 64-pole in the restricted range in our analysis (Case 3, varied foreground model with foreground effects included in the covariance matrix), we obtained unbiased results for all cosmological parameters up to $k_{\rm max} = 0.08 \,h\,\text{Mpc}^{-1}$, a slightly more restricted range than we find for the hexadecapole. This is likely due to how the foreground removal effect suppresses our cosmological signal in the covariance matrix, thus decreasing the error budget, combined with the 64-pole being highly non-linear.

Removing the foreground effect from the covariance matrix yields a much larger error budget, and we tried including the 64-pole in this case. We found that indeed we obtain unbiased results up to $k_{\rm max} = 0.09 \,h\,\text{Mpc}^{-1}$ in this case but with very large uncertainties on our parameters, as seen in \autoref{MCMCtable_errors_P6}.

Our results show that in the absence of foregrounds, the 64-pole can improve constraints without biasing parameter estimates. In the foreground removed case, the 64-pole does not improve constraints, but the 64-pole could still be useful in analysing foreground cleaned IM data. This is because its underlying cosmological signal is quite weak, but it is highly sensitive to the effects of the beam and foreground removal, or other unidentified systematics. It could thus be used as a further check for any residual systematic effects that might be present in the data.

\section{Conclusions}\label{sec:conclusion}

The aim of this work was to perform a comprehensive cosmological parameter estimation with the \hi IM power spectrum multipoles, and investigate the level of uncertainties future surveys like the SKA can realistically obtain, requiring unbiased estimates. We used modelling and simulations of \hi IM that account for effects of the telescope beam and foreground removal, and performed MCMC analyses on these. We also showed how the beam and foreground removal effects impact the covariance matrix and higher order multipoles. We summarise our main findings and conclusions below:

\begin{itemize}[leftmargin=*]
    \item In the absence of foregrounds, we are able to retrieve unbiased estimates for cosmological parameters using our model, with below 10\% percent level uncertainties (and for the transverse AP parameter, below 1\% percent level uncertainty). Including the hexadecapole in our analysis does not bias parameter estimates if we only consider it at a restricted range. Even at a restricted range, including the hexadecapole significantly decreases parameter uncertainties for all cases considered. In particular, when including the restricted hexadecapole, we are able to retrieve the growth rate parameter ($\overline{T}_\hinospace f\sigma_8$) with 8.8\% uncertainty.
    \item In the presence of a telescope beam and foreground removal effects, it is crucial to include the modelling of these in the covariance matrix as it makes a significant difference. In particular, the covariance matrices between different multipoles become non-negligible as these effects change the correlations between multipoles.
    \item If we do not account for the effects of foreground removal in the modelling, we obtain significantly biased parameter estimates (see also the very recent study by \citet{Cunnington:2020wdu} for the case of primordial non-gaussianity measurements).
    \item We therefore develop a 2-parameter foreground model to account for the removal of modes that occurs due to foreground cleaning. With no assumptions about the foreground removal process (i.e. by letting these parameters vary), we use this model to try and recover unbiased cosmological parameter estimates and succeed, finding that the two extra free parameters are enough to model the effects of foreground removal in our case.
    \item We find that we are able to model the effects of foreground removal, and recover the growth rate parameter ($\overline{T}_\hinospace f\sigma_8$) uncertainty to be 13.3\%, slightly larger than in the foreground-free case. The other cosmological parameters also experience a slight increase in uncertainties, but they are not as significant.
    \item We investigate the effect of including the 64-pole in our analysis. We find that for the foreground-free case, it improves parameter uncertainties without biasing them, but worsens constraints in the foreground removed case. However, we propose that the 64-pole could be a useful tool to investigate systematic effects in foreground cleaned intensity mapping data, since it is highly sensitive to these.
\end{itemize}

The results in this paper are dependent on our choice of simulation and noise modelling. It would be interesting in future work to test the robustness of our modelling against more complex foreground simulations, for example including polarization leakage. Considering additional noise and systematic effects, such as (1/$f$) noise and RFI flagging in our simulations would also be worthwhile.

Although we looked specifically at the single dish \hi intensity mapping case in this paper, our findings could also be relevant for the interferometer case. The interferometer case would have much better angular resolution, but additional complications due to different instrumental systematic effects. However, foreground removal effects are analogous and equally important for both cases, and \fastica\ in particular has been applied to interferometric data already (see e.g. \citet{Chapman:2012yj,Hothi:2020dgq}).

To further investigate our results for the covariance matrix, future plans include calculating the covariance matrix using a suite of simulations, and obtaining a more robust estimate of how systematic effects impact the \hi IM covariance matrix.

We hope our findings can be useful for analysing \hi intensity mapping data from the MeerKAT single-dish survey \citep{Santos:2017qgq, Pourtsidou:2017era, Li:2020bcr, wang2020hi}, in particular by using multipole expansion and our modelling prescriptions for understanding systematic effects. Our formalism can also help the preparation of forthcoming observations by providing realistic forecasts.

\section*{Acknowledgements}
We are grateful to the anonymous reviewer for very useful comments and suggestions that improved the quality of this paper. We thank Seshadri Nadathur, Mario Santos and Marta Spinelli for useful discussions and feedback. PS is supported by the Science and Technology Facilities Council [grant number ST/P006760/1] through the DISCnet Centre for Doctoral Training. SC is supported by STFC grant ST/S000437/1. AP is a UK Research and Innovation Future Leaders Fellow, grant MR/S016066/1, and also acknowledges support by STFC grant ST/S000437/1. This research utilised Queen Mary’s Apocrita HPC facility, supported by QMUL Research-IT \url{http://doi.org/10.5281/zenodo.438045}. We acknowledge the use of open source software \citep{scipy:2001,Hunter:2007,  mckinney-proc-scipy-2010, numpy:2011,  Lewis1999bs,Lewis2019xzd}. Some of the results in this paper have been derived using the \texttt{healpy} and \texttt{HEALPix} package. We thank New Mexico State University (USA) and Instituto de Astrofisica de Andalucia CSIC (Spain) for hosting the Skies \& Universes site for cosmological simulation products. 

\section*{Data Availability}

The data underlying this article will be shared on reasonable request to the corresponding author.




\bibliographystyle{mnras}
\bibliography{Bib} 




\appendix

\section{Covariance matrix in the presence of a telescope beam}\label{CovAppendix}

In order to better understand why the telescope beam is increasing correlations between different multipoles, we consider a toy power spectrum model with and without the telescope beam effect. 

We begin with the case of no telescope beam. First, assume we have a simple, isotropic matter power spectrum (no RSD): $P_\hinospace(k) = \overline{T}_\hinospace^2 \, b_\hinospace^2 P_{\rm m}(k)$. Let us also set $P_N = 0$, $b_\hinospace = 1$ and $\overline{T}_\hinospace = 1$ to obtain $P_\hinospace(k) = P_{\rm m}(k)$. This yields  $\sigma^{2} (k, \mu) = \sigma^{2} (k) = P_{\rm m}^{2}(k) / N_{\rm modes}(k)$. The sub-covariance matrices become:
\begin{equation}\label{covLLprime}
    C_{\ell\ell^\prime}(k) = \frac{(2\ell + 1)(2\ell^\prime + 1)}{2} \frac{P_{\rm m}^{2}(k)}{N_{\rm modes}(k)} \int^{1}_{-1} d\mu \, \mathcal{L}_{\ell}(\mu) \mathcal{L}_{\ell^\prime}(\mu) \, .
\end{equation}
In the absence of RSD and any other anisotropic effect in the power spectrum, the quadrupole and hexadecapole are null. Using \autoref{covLLprime} we can confirm that the off-diagonal covariance matrix terms that include these multipoles are also null, as expected: $C_{02}(k) = C_{04}(k) = C_{24}(k)=0$.

Next we consider the same power spectrum ($P_N = 0$, $b_\hinospace = 1$ and $\overline{T}_\hinospace = 1$) but with a telescope beam damping term, such that $P_\hinospace(k, \mu) = P_{\rm m}(k) \widetilde{B}^2_\perp(k,\mu) = P_{\rm m}(k) \exp\left(-k^2 R_\text{beam}^2(1-\mu^2)\right)$. We assume $R_\text{beam} = 1\,\text{Mpc}\,h^{-1}$ for simplicity. This yields a covariance per $k$ and $\mu$ bin of: 
\begin{equation}
    \sigma^{2} (k, \mu) = \frac{P_{\rm m}^{2}(k)  \exp\left(-2k^2 (1-\mu^2)\right)}{N_{\rm modes}(k)}\, ,
\end{equation}
and sub-covariance matrices given by:
\begin{equation}
\begin{split}
    C_{\ell\ell^\prime}(k) = \ & \frac{(2\ell + 1)(2\ell^\prime + 1)}{2} \frac{P_{\rm m}^{2}(k)}{N_{\rm modes}(k)} \\
    \ & \times \int^{1}_{-1} d\mu \, e^{-2k^2 (1-\mu^2)} \mathcal{L}_{\ell}(\mu) \mathcal{L}_{\ell^\prime}(\mu) \, ,
\end{split}
\end{equation}
yielding the following off-diagonal covariance matrices:
\begin{equation}
\begin{split}
    C_{02}(k) \ & = \frac{5}{2} \frac{P_{\rm m}^{2}(k)}{N_{\rm modes}(k)}  \int^{1}_{-1} d\mu \, \mathcal{L}_{0}(\mu) \mathcal{L}_{2}(\mu) e^{-2k^2 (1-\mu^2)} \\ 
    \ & = \frac{P_{\rm m}^{2}(k)}{N_{\rm modes}(k)} \cdot \frac{ 60k - 5\sqrt{2\pi}e^{-2k^2}(4k^2 + 3) \text{erfi}(\sqrt{2}k) }{32k^3} \, ,
\end{split}
\end{equation}  
\begin{equation}
\begin{split}
    C_{04}(k) \ & = \frac{9}{2} \frac{P_{\rm m}^{2}(k)}{N_{\rm modes}(k)}  \int^{1}_{-1} d\mu \, \mathcal{L}_{0}(\mu) \mathcal{L}_{4}(\mu) e^{-2k^2 (1-\mu^2)} \\
    \ & = \frac{P_{\rm m}^{2}(k)}{N_{\rm modes}(k)} \cdot \frac{9}{512 k^5} [3\sqrt{2\pi}e^{-2k^2} (16k^4 + 40k^2 + 35) \\ 
    \ & \times \text{erfi}(\sqrt{2}k) +20(4k^3 - 21k)] \, ,
\end{split}
\end{equation}
\begin{equation}
\begin{split}
    C_{24}(k) \ & = \frac{45}{2} \frac{P_{\rm m}^{2}(k)}{N_{\rm modes}(k)} \int^{1}_{-1} d\mu \, \mathcal{L}_{2}(\mu) \mathcal{L}_{4}(\mu) e^{-2k^2 (1-\mu^2)} \\ 
    \ & = \frac{P_{\rm m}^{2}(k)}{N_{\rm modes}(k)} \cdot \frac{45}{4096 k^7} [4(304k^5 - 600k^3 + 1575k) \\ 
    \ & - 3\sqrt{2\pi}e^{-2k^2}(64k^6 + 208k^4 + 500k^2 + 525)\text{erfi}(\sqrt{2}k)]\, ,
\end{split}
\end{equation}
where $\text{erfi}(x)$ is the imaginary error function. From the results above, we can see that even in the absence of RSD, when we include the effect of a telescope beam the power spectrum does not have a vanishing covariance between the different multipoles.



\bsp	
\label{lastpage}
\end{document}